\documentclass[aip,graphicx]{revtex4-1}
\usepackage{graphicx}
\usepackage{amsfonts}
\usepackage{amsmath}
\usepackage{physics}
\usepackage{xcolor}

\def\Xint#1{\mathchoice
{\XXint\displaystyle\textstyle{#1}}%
{\XXint\textstyle\scriptstyle{#1}}%
{\XXint\scriptstyle\scriptscriptstyle{#1}}%
{\XXint\scriptscriptstyle\scriptscriptstyle{#1}}%
\!\int}
\def\XXint#1#2#3{{\setbox0=\hbox{$#1{#2#3}{\int}$}
\vcenter{\hbox{$#2#3$}}\kern-.5\wd0}}

\def\dashint{\Xint-}

\draft 

\begin{document}


\title{A general framework for the study of electrostatic point charges in multilayer planar structures} 


\author{George Fikioris}%
\email{gfiki@ece.ntua.gr}
\affiliation{School of Electrical and Computer Engineering, National Technical University of Athens, 9 Iroon Polytechneiou, Athens 15780, Greece.}%

\author{Theodoros T. Koutserimpas}
\email{tkoutserimpas@mail.ntua.gr}
\affiliation{School of Electrical and Computer Engineering, National Technical University of Athens, 9 Iroon Polytechneiou, Athens 15780, Greece.}%

\author{Elias N. Glytsis}%
\email{eglytsis@central.ntua.gr}
\affiliation{School of Electrical and Computer Engineering, National Technical University of Athens, 9 Iroon Polytechneiou, Athens 15780, Greece.}%


\begin{abstract}
We develop a general framework for the electrostatic analysis of point charges in multilayer planar structures with arbitrary layer thicknesses and material parameters. Starting from a Hankel-transform analysis, we derive alternative representations of the solution and establish a Stokes-like formulation based on ``generalized reflection coefficients,'' yielding a systematic and physically transparent treatment of multilayer media. This approach extends classical image theory to parameter regimes in which the conventional image-charge series (which has an infinite number of terms) diverges. The formulation applies to arbitrary permittivity values, including negative permittivities, where overscreening effects and plasmon-resonant conditions may occur. In these regimes, we show that the boundary-value problem no longer has a unique solution because homogeneous (source-free) modes appear; and we derive Cauchy-principal-value integral representations for the particular solution. We also introduce an asymptotic ``phantom-image'' method that replaces a divergent infinite image series by a finite set of effective sources, thus providing a computationally efficient approximation in large-reflection regimes. These results furnish both practical computational tools and additional mathematical insight into the structure of electrostatic image theory in layered media.

\end{abstract}


\maketitle 

\section{\label{sec:intro} Introduction}
Image theory is a fundamental concept taught in introductory courses on physics and electromagnetism, providing a robust framework for developing physical intuition about field solutions under geometric symmetries. Solutions to the Laplace equation are presented rigorously, with both mathematical and physical significance. \cite{Griffiths1, Wait1} In image theory, a series of image charges are introduced in a hypothetical homogeneous medium positioned outside the computational region of interest, to satisfy the appropriate boundary conditions. In standard dielectric configurations, the resulting image series is typically convergent.

For planar layered media, one of the most familiar tools in physics is the transfer-matrix approach, which has long been used in stratified wave-propagation and optical multilayer problems. In that approach, the field amplitudes in adjacent layers are related through interface and propagation matrices, and the overall response is obtained by multiplying the corresponding matrices layer by layer. This framework is especially natural for wave problems, where propagation through a finite thickness and successive reflections/transmissions are the central physical ingredients. \cite{Chew1, Paul1, Bellman1, Yeh1, Mackay1} Closely related ideas also appear in invariant-embedding and generalized-reflection formulations. \cite{Bellman1, Chew1} For electrostatic multilayers, however, the transfer-matrix viewpoint is not always the most transparent route to closed-form image-theoretic representations, to the analysis of divergent image series, or to the explicit construction of homogeneous resonant solutions. In the present work, we therefore adopt a Hankel-transform formulation together with Stokes-like generalized reflection coefficients. Besides leading to a physically transparent recursive treatment of multilayers, this formulation is particularly advantageous for the analysis developed later in the paper, especially in Section~\ref{sec:homogeneous}, where it naturally yields homogeneous source-free solutions for an arbitrary number of layers.

Previous studies on calculating the electrostatic potential in three-dielectric media with planar interfaces have demonstrated the connection between classical image theory and the Hankel transform.\cite{Barrera, Barrera_corrections, Rahman} The Hankel transform has also been employed to calculate electrostatic potentials in Cold Atmospheric-Pressure Plasma Jets.\cite{Vafeas} Furthermore, a generic solution for $N$ layers of dielectrics with equal spacing has been derived using classical image theory. \cite{Wang} Despite these advancements, a comprehensive and systematic approach for calculating the electrostatic potential in multilayer planar structures with arbitrary thicknesses and arbitrary material parameters, including parameter regimes with negative permittivities and possible resonant singularities, remains absent in the literature.

Assuming the long-wavelength limit, an external perturbation may induce an electric displacement field, $D$, only for $k \sim L^{-1}\to 0$, where $L$ is the macroscopic size of the system. This makes spatial dispersion negligible. In such cases, the Kramers--Kronig relation and therefore causality imply that negative static permittivity is only possible for active media, as expressed by the sum rule\cite{Sanders,Chiao1,Chiao2}
\begin{equation}
\varepsilon (0) = 1 + \frac{2}{\pi }\dashint_0^\infty  d\omega \,\frac{{\rm Im}\,\varepsilon (\omega )}{\omega } < 0.
\end{equation}
Notably, negative longitudinal static permittivity may also be permitted for passive systems with positive inelastic electron scattering when spatial dispersion is taken into account.\cite{Dolgov,Aniya,Manolescu,Nazarov}

The difficulties of image theory in sign-indefinite permittivity regimes have been appreciated in other geometries as well. In particular, Majic studied the problem of a point charge outside a dielectric sphere with negative permittivity and showed that the standard image representation ceases to be straightforward in the resonant regime; in that case, the classical image construction must be corrected in order to treat divergences associated with the spherical image representation.\cite{Majic1} That work highlighted, in a different geometry, that the extension of image theory to negative-permittivity media is subtle and may require a careful treatment of divergent representations. By contrast, in the present planar multilayer problem, our formulation is built throughout on Hankel-type integral representations, which are convergent whenever the boundary-value problem is in the regular regime. Even in parameter ranges where the conventional image series diverges, our starting point is not an \emph{ad hoc} assignment of values to a divergent series, but rather a transformed-domain representation whose continuation and principal-value interpretation arise directly from the boundary-value problem itself. In this sense, when our integral formulas can be viewed \emph{a posteriori} as regularizing divergent image series, the regularization is not introduced arbitrarily, but is inherited from the analytically continued integral solution.

Beyond extreme cases involving active media, the study of electrostatic fields and potentials in negative-permittivity environments is also important for plasmonic structures. Such structures exhibit negative permittivities below their plasma frequency, and in the quasistatic limit may support source-free resonant states. \cite{Mayergoyz1,Fourn1, Jung1,Mayergoyz-book} Image-theory constructions in planar structures are also relevant to the quasistatic analysis of Veselago-type lenses and related layered systems with sign-changing constitutive parameters. \cite{Farhi1} In the present work, the resonant regime is identified in a precise mathematical manner: It corresponds to the parameter region in which the transformed-domain denominator develops a pole on the positive real axis, so that a homogeneous source-free mode appears. In the three-layer and grounded-layer problems studied here, this is precisely the regime that we later term the non-unique/divergent-series (NU-DS) region. Thus, rather than invoking ``plasmon resonant conditions'' only heuristically, our analysis provides an explicit criterion for when such quasistatic plasmon-like resonant behavior occurs in layered planar media.

The general framework developed in this paper starts from a Hankel-transform formulation, used to derive Stokes-like generalized reflection coefficients that provide a systematic recursive treatment of the multilayer problem. This viewpoint preserves the physical intuition of successive reflections and transmissions while remaining better suited than the standard transfer-matrix picture to the explicit construction of image-theoretic representations and homogeneous resonant solutions. We then show how different parameter regimes lead to qualitatively different mathematical structures: convergent image series, convergent integrals with divergent image series, and non-unique solutions associated with poles and source-free modes. For resonant cases we derive principal-value representations for the particular solution, and we also introduce a phantom-image method that replaces a divergent infinite image series by a finite set of effective sources in appropriate asymptotic regimes.

\section{\label{sec:sec_ii} Hankel-Transform Solution for Multilayer Structures}

We begin by giving some well-known properties of the zero-order Hankel transform (HT). \cite{Piessens1, Lebedev1} Let $\rho > 0$. The HT $\bar g(k)$ of a function $g(\rho )$ is defined by
\begin{equation}
\mathcal{H}\left\{ {g(\rho );k} \right\} = \bar g(k) = \int\limits_0^\infty  {g(\rho ){J_0}(k\rho )\rho {\kern 1pt} d\rho ,\quad k > 0} 
\end{equation}
\noindent where ${J_0}$ is the zero-order Bessel function of the first kind. The inversion formula is 
\begin{equation}
{\mathcal{H}^{ - 1}}\left\{ {\bar g(k);\rho } \right\} = g(\rho ) = \int\limits_0^\infty  {\bar g(k){J_0}(k\rho )k{\kern 1pt} dk,\quad \rho  > 0} .
\label{eq:inv1}
\end{equation}
The following inverse Hankel transform
\begin{equation}
{\mathcal{H}^{ - 1}}\left\{ {\frac{{{e^{ - k\left| z \right|}}}}{k};\rho } \right\} = \int_0^\infty  {{J_0}(k\rho ){e^{ - k\left| z \right|}}dk = \frac{1}{{\sqrt {{\rho ^2} + {z^2}} }}} ,\quad \rho  > 0,\quad z \in \mathbb{R}
\label{eq:inv2}
\end{equation}
\noindent will be used throughout. Now let $(\rho ,\varphi ,z)$ denote the cylindrical coordinates. The Laplacian $\Delta $ of a $\varphi$-independent function $f(\rho ,z)$ is 
\begin{equation}
\Delta f(\rho ,z) = \frac{1}{\rho }\frac{\partial }{{\partial \rho }}\left[ {\rho \frac{{\partial f(\rho ,z)}}{{\partial \rho }}} \right] + \frac{{{\partial ^2}f(\rho ,z)}}{{\partial {z^2}}}.
\end{equation}
Using elementary Bessel-function properties, we find that the HT of $\Delta f(\rho ,z)=0$ (with respect to $\rho$) is

\begin{equation}
\mathcal{H}\left\{ {\Delta f(\rho ,z);k} \right\} =  - {k^2}\bar f(k,z) + \frac{{{\partial ^2}\bar f(k,z)}}{{\partial {z^2}}}=0,
\end{equation}
\noindent whose two linearly independent solutions are $\bar f(k,z) = {e^{ \pm kz}}$. We have thus found that the general solution to the $\varphi$-independent Laplace equation $\Delta f(\rho ,z) = 0$ is
\begin{equation}
f(\rho ,z) = \int\limits_0^\infty  {\left[ {\alpha(k){e^{ - kz}} + \beta(k){e^{kz}}} \right]{J_0}(k\rho ){\kern 1pt} dk}. 
\label{eq:gen}
\end{equation}
\noindent This result (which is often arrived at using separation of variables \cite{Wait1}) holds because $f(\rho,z)=e^{\pm kz} J_0(k\rho)$ satisfies $\Delta f(\rho ,z) = 0$ for all $k>0$. Note that the quantity inside the square brackets equals $k\bar f(k,z)$. When applying (\ref{eq:gen}) to Poisson boundary value problems, one should anticipate $\beta(k)$ and $\alpha(k)$ to turn out exponentially small (as $k\to +\infty$) for $z>0$ and $z<0$, respectively. 

\begin{figure}
\includegraphics{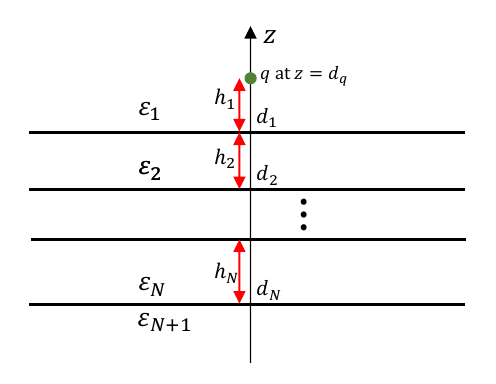}
\caption{\label{fig:fig_1} Multilayer planar structure with electric charge $q$ located at $z = {d_q}$ in the top layer. The top and bottom layers, which are numbered 1 and $N+1$, extend to infinity}
\end{figure}

Our multilayer structure is planar, see Fig.~\ref{fig:fig_1}, with layers numbered $1,2, \ldots ,N + 1$, where $N \ge 1$. Layer $n$ has permittivity ${\varepsilon _n}$ (with $\varepsilon_n\in\mathbb{R}$) and its lower boundary at $z = {d_n}$. Layers $1$ and $N+1$ are half-spaces, extending to $z =  + \infty $ and $z =  - \infty $, respectively. A point charge $q$ is located at $z = {d_q}$ in the top layer 1. Thus
\begin{equation}
{d_q} > {d_1} > {d_2} >  \ldots  > {d_N}.
\label{eq:d}
\end{equation}
We occasionally denote the thickness of the $n$th layer by $h_n$ (see Fig.~\ref{fig:fig_1}), so that
\begin{equation}
\label{eq:h}
h_1=d_q-d_1,\quad
    h_n=d_{n-1}-d_n,\quad n=2,3,\ldots,N.
\end{equation}

Let ${V_n}(\rho ,z)$ be the $\varphi$-independent potential in layer $n$ and let $\delta $ be the Dirac delta function. Then the following Poisson/Laplace equations and boundary conditions hold,
\begin{eqnarray}
{\Delta {V_1}(\rho ,z) =  - {{q\delta (x)\delta (y)\delta (z - {d_q})} \mathord{\left/
 {\vphantom {{q\delta (x)\delta (y)\delta (z - {d_q})} {{\varepsilon _1}}}} \right.
 \kern-\nulldelimiterspace} {{\varepsilon _1}}}},
 \label{eq:Poisson}
 \\
\Delta {V_n}(\rho ,z) = 0,\quad n = 2, \ldots ,N + 1,
\label{eq:Laplace}
\\
{V_n}(\rho ,d_n^ + ) = {V_{n + 1}}(\rho ,d_n^ - ),\quad n = 1,2, \ldots ,N,
\label{eq:bc1}
\\
{\varepsilon _n}{\partial _z}{V_n}(\rho ,d_n^ + ) = {\varepsilon _{n + 1}}{\partial _z}{V_{n + 1}}(\rho ,d_n^ - ),\quad n = 1,2, \ldots ,N,
\label{eq:bc2}
\\
{V_1}(\rho , + \infty ) = 0,
\label{eq:bcinfp}
\\
{V_{N + 1}}(\rho , - \infty ) = 0.
\label{eq:bcinfm}
\end{eqnarray}
The solution to the Laplace equation~(\ref{eq:Laplace}) is given by~(\ref{eq:gen}). The solution to the Poisson equation~(\ref{eq:Poisson}) requires an additional term, namely the primal potential due to the point charge $q$. Thus the general solutions to~(\ref{eq:Poisson}) and~(\ref{eq:Laplace}) can be written as
\begin{equation}
\begin{aligned}
{V_n}\left( {\rho ,z} \right) = \frac{q}{{4\pi {\varepsilon _n}}}\left( {\frac{{{\delta _{1n}}}}{{\sqrt {{\rho ^2} + {{(z - {d_q})}^2}} }} + \int_0^\infty  {{J_0}(k\rho )\left[ {{A_n}(k){e^{ - kz}} + {B_n}(k){e^{kz}}} \right]dk} } \right),\\
\;n = 1, \ldots ,N + 1,
\end{aligned}
\label{eq:gensol}
\end{equation}
\noindent where ${\delta _{1n}}$ is the Kronecker delta. We will find cases where ${A_n}(k)$ and/or ${B_n}(k)$ have a simple pole at a point $k = {k_0} > 0$ (with ${k_0}$ to be determined). When simple poles do turn up (depending on the values of ${\varepsilon _n}$), the integral in~(\ref{eq:gensol}) (as well as all integrals that follow) must be properly interpreted. This modifies the conventional HT-procedure,  as will be first discussed in Section \ref{sec:sec_vii}. 

Use of~(\ref{eq:inv2}) allows us to incorporate the primal potential into the integral, giving the alternative to~(\ref{eq:gensol}) equation
\begin{equation}
{V_n}\left( {\rho ,z} \right) = \frac{q}{{4\pi {\varepsilon _n}}}\int_0^\infty  {{J_0}(k\rho )\left[ {{\delta _{1n}}{e^{ - k\left| {z - {d_q}} \right|}} + {A_n}(k){e^{ - kz}} + {B_n}(k){e^{kz}}} \right]dk} ,\quad n = 1, \ldots ,N + 1,
\label{eq:gensolalt}
\end{equation}
\noindent which, by~(\ref{eq:inv1}), means that the HT ${\overline V _n}\left( {k,z} \right)$ of ${V_n}\left( {\rho ,z} \right)$ can be found from
\begin{equation}
k{\kern 1pt} {\overline V _n}\left( {k,z} \right) = \frac{q}{{4\pi {\varepsilon _n}}}\left[ {{\delta _{1n}}{e^{ - k\left| {z - {d_q}} \right|}} + {A_n}(k){e^{ - kz}} + {B_n}(k){e^{kz}}} \right],\quad n = 1, \ldots ,N + 1.
\label{eq:gensolaltk}
\end{equation}
The boundary conditions~(\ref{eq:bc1})--(\ref{eq:bcinfm}) continue to hold if the several ${V_n}\left( {\rho ,d} \right)$ are replaced by their HTs ${\overline V _n}\left( {k,d} \right)$. Thus~(\ref{eq:gensolaltk}), (\ref{eq:bcinfp}) and (\ref{eq:bcinfm}) immediately give
\begin{eqnarray}
{B_1}(k) = 0
\label{eq:B0},
 \\
{A_{N + 1}}(k) = 0.
\label{eq:A0}
\end{eqnarray}

To apply the remaining conditions (\ref{eq:bc1}) and (\ref{eq:bc2}) we define the dimensionless quantities
\begin{equation}
{R_{nm}} = \frac{{{\varepsilon _n} - {\varepsilon _m}}}{{{\varepsilon _n} + {\varepsilon _m}}},\quad {T_{nm}}  = \frac{{2{\varepsilon _m}}}{{{\varepsilon _n} + {\varepsilon _m}}},\quad 1 \le n,m \le N + 1.
\label{eq:coef}
\end{equation}
We call ${R_{nm}}$ and ${T_{nm}}$ the reflection and transmission coefficients, respectively, as they are directly analogous to familiar quantities of wave propagation and transmission line problems. \cite{Chew1, Paul1, Bellman1} The equations
\begin{equation}
- {R_{mn}} = {R_{nm}} = 1 - {T_{nm}},\quad {T_{nm}} = \frac{{{\varepsilon _m}}}{{{\varepsilon _n}}}{T_{mn}}
\label{eq:coef2},
\end{equation}
which we use throughout (sometimes without special mention) follow from our definitions. Some algebra now shows that (\ref{eq:d}), (\ref{eq:bc1}), (\ref{eq:bc2}), (\ref{eq:gensolaltk}) and (\ref{eq:coef}) yield
\begin{eqnarray}
{A_n}\left( k \right) = \frac{1}{{{T_{n,n + 1}}}}\left[ {{A_{n + 1}}\left( k \right) + {B_{n + 1}}\left( k \right){R_{n,n + 1}}{e^{2k{d_n}}}} \right],\quad n = 1,2,3, \ldots ,N,
\label{eq:a}
 \\
{B_n}\left( k \right) = \frac{1}{{{T_{n,n + 1}}}}\left[ {{A_{n + 1}}\left( k \right){R_{n,n + 1}}{e^{ - 2k{d_n}}} + {B_{n + 1}}\left( k \right)} \right],\quad n = 2,3, \ldots ,N,
\label{eq:b}
\\
{e^{ - k{d_q}}} = \frac{1}{{{T_{12}}}}\left[ {{A_2}\left( k \right){R_{12}}{e^{ - 2k{d_1}}} + {B_2}\left( k \right)} \right]
\label{eq:exp}.
\end{eqnarray}
Note that (\ref{eq:a}) holds for $n = 1$, but that (\ref{eq:b}) does not. Since ${R_{n,n + 1}}$ and ${T_{n,n + 1}}$ are known, (\ref{eq:B0}), (\ref{eq:A0}), and (\ref{eq:a})--(\ref{eq:exp}) make up a $(2N + 2) \times (2N + 2)$ system of linear algebraic equations for the $k$-dependent coefficients ${A_1},{A_2}, \ldots ,{A_{N + 1}},{B_1},{B_2}, \ldots ,{B_{N + 1}}$. Once we solve the system, the $n$th equation in (\ref{eq:gensol}) or (\ref{eq:gensolalt}) can be regarded as an integral representation of the potential ${V_n}(\rho ,z)$ in any layer $n$. An analytical solution of the system is possible for any given small value of $N$, especially if one uses matrices and manipulates them symbolically. The next section discusses an alternative method.

\section{\label{sec:sec_iii} Stokes-like generalized reflection coefficients}

Even when we can blindly solve the aforementioned $(2N + 2) \times (2N + 2)$ system, it is easier algebraically---and more transparent physically and analytically---to work with generalized reflection coefficients; these are wave-propagation concepts that we will now adapt to the present (electrostatics) problem. Generalized reflection coefficients were first used by Stokes in 1883, in a study \cite{Stokes1} of the reflection and transmission of light waves impinging upon a stack of glass plates. Stokes used physical reasoning to derive recurrence relations for his coefficients, and this reasoning is followed reasonably closely in the monograph of Bellman and Wing. \cite{Bellman1} The textbook by Chew \cite{Chew1} (see also the references therein) provides similar relations for the case of electromagnetic plane-wave propagation in layered media. Chew \cite{Chew1}  also describes how such coefficients can be applied to other cases, including non-layered inhomogenous problems in which a dielectric constant varies continuously. 

In this section, we define such coefficients for the present problem and, using straightforward analytical manipulations, find the analogous recurrence relations that our coefficients obey. Then we give an ensuing, step-by-step procedure that enables a more straightforward determination of ${A_1},{A_2}, \ldots ,{A_{N + 1}},{B_1},{B_2}, \ldots ,{B_{N + 1}}$. Our notation is analogous to that of Chew.\cite{Chew1} Thus, our generalized reflection coefficients are denoted by ${\tilde R_{n,n + 1}}$. As opposed to the various wave-propagation problems in layered media, however, our ${\tilde R_{n,n + 1}}$ depend on the HT variable $k$.

For $n = 1,2, \ldots ,N$, (\ref{eq:B0}) tells us that the ${\overline V _n}\left( {k,z} \right)$ of (\ref{eq:gensolaltk}) has exactly two terms. We define the generalized reflection coefficient ${\tilde R_{n,n + 1}}\left( k \right)$ to be the ratio of those terms evaluated at $z = {d_n}$ (i.e., at the boundary that separates layers $n$ and $n+1$), according to
\begin{equation}
{\tilde R_{n,n + 1}}\left( k \right) =\begin{cases}
{{A_1}(k){e^{ - k(2{d_1} - {d_q})}},\quad n = 1,}\\
{\frac{{{A_n}(k){e^{ - 2k{d_n}}}}}{{{B_n}(k)}},{\kern 1pt} \quad n = 2,3, \ldots ,N,}
\end{cases}
\label{eq:R1}
\end{equation}
\noindent where, for the case $n=1$, we used (\ref{eq:B0}). Now divide (\ref{eq:a}) by (\ref{eq:b})---or, for the case $n=1$ by (\ref{eq:exp})---and then use (\ref{eq:R1}) to obtain
\begin{equation}
{\tilde R_{n,n + 1}}\left( k \right) = \frac{{{A_{n + 1}}(k){e^{ - 2k{d_n}}} + {R_{n,n + 1}}{B_{n + 1}}(k)}}{{{R_{n,n + 1}}{A_{n + 1}}(k){e^{ - 2k{d_n}}} + {B_{n + 1}}(k)}},{\kern 1pt} \quad n = 1,2,3, \ldots ,N
\label{eq:R2}.
\end{equation}
For $n=N$, (\ref{eq:R2}) and (\ref{eq:A0}) give
\begin{equation}
{\tilde R_{N,N + 1}}\left( k \right) = {R_{N,N + 1}}
\label{eq:R3}.
\end{equation}
If $N=1$ (two-layer problem), there is only one ${\tilde R_{n,n + 1}}\left( k \right)$, namely ${\tilde R_{1,2}}\left( k \right)$, and it is found by putting $N=1$ in (\ref{eq:R3}). For $N \ne 1$ and $n \ne N$, divide the numerator and denominator of (\ref{eq:R2}) by the nonzero ${B_{n + 1}}(k)$. Then, use (\ref{eq:R1}) to express the ratio ${A_{n + 1}}(k)/{B_{n + 1}}(k)$ in terms of ${\tilde R_{n + 1,n + 2}}\left( k \right)$. The result is
\begin{equation}
{\tilde R_{n,n + 1}}\left( k \right) = \frac{{{R_{n,n + 1}} + {{\tilde R}_{n + 1,n + 2}}\left( k \right){e^{ - 2k\left( {{d_n} - {d_{n + 1}}} \right)}}}}{{1 + {R_{n,n + 1}}{{\tilde R}_{n + 1,n + 2}}\left( k \right){e^{ - 2k\left( {{d_n} - {d_{n + 1}}} \right)}}}},{\kern 1pt} \quad n = 1,2,3, \ldots ,N - 1
\label{eq:R4}.
\end{equation}
Eq.~(\ref{eq:R4}), which is an analogue of  Eq. (2.1.24) of Chew,\cite{Chew1} is the desired Stokes-like recurrence relation for our ${\tilde R_{n,n + 1}}\left( k \right)$. An alternative follows directly from (\ref{eq:R4}) and (\ref{eq:coef2}):
\begin{equation}
\begin{aligned}
{\tilde R_{n,n + 1}}\left( k \right) = {R_{n,n + 1}} + \frac{{{T_{n,n + 1}}{T_{n + 1,n}}{{\tilde R}_{n + 1,n + 2}}\left( k \right){e^{ - 2k\left( {{d_n} - {d_{n + 1}}} \right)}}}}{{1 + {R_{n,n + 1}}{{\tilde R}_{n + 1,n + 2}}\left( k \right){e^{ - 2k\left( {{d_n} - {d_{n + 1}}} \right)}}}},{\kern 1pt} \\
\quad n = 1,2,3,\ldots ,N - 1.
\end{aligned}
\label{eq:R5}
\end{equation}
Eq. (\ref{eq:R5}) is an analogue of Chew's\cite{Chew1} Eq. (2.1.23) and  explicitly gives the difference ${\tilde R_{n,n + 1}}\left( k \right) - {R_{n,n + 1}}$.

Suppose that $N \ne 1$. For $n = 2,3, \ldots ,N$, write (\ref{eq:a}) with $n-1$ in place of $n$ to get
\begin{equation}
{A_{n - 1}}\left( k \right) = \frac{{{A_n}\left( k \right)}}{{{T_{n - 1,n}}}}\left[ {1 + \frac{{{B_n}\left( k \right)}}{{{A_n}\left( k \right)}}{R_{n - 1,n}}{e^{2k{d_{n - 1}}}}} \right],\quad n = 2,3, \ldots ,N
\label{eq:R6}.
\end{equation}
Upon replacing the ratio ${B_n}(k)/{A_n}(k)$ from the bottom equation (\ref{eq:R1}) and solving for ${A_n}\left( k \right)$, we obtain
\begin{equation}
{A_n}\left( k \right) = \frac{{{T_{n - 1,n}}{A_{n - 1}}\left( k \right){{\tilde R}_{n,n + 1}}\left( k \right)}}{{{{\tilde R}_{n,n + 1}}\left( k \right) + {R_{n - 1,n}}{e^{ - 2k\left( {{d_n} - {d_{n - 1}}} \right)}}}},\quad n = 2,3, \ldots ,N
\label{eq:R7},
\end{equation}
\noindent an equation that is somewhat analogous to Chew’s \cite{Chew1} equation (2.1.26). 

The above-derived formulas give rise to the following algorithm for the analytical (and, for many layers, symbolical) calculation of ${A_1}\left( k \right), \ldots ,{A_{N + 1}}\left( k \right),{B_1}\left( k \right), \ldots ,{B_{N + 1}}\left( k \right)$.
\begin{enumerate}
\item Using the initial value (\ref{eq:R3}), apply (\ref{eq:R4}) (or its alternative (\ref{eq:R5})) successively for $n = N - 1,N - 2, \ldots ,1$, thus obtaining all ${\tilde R_{N,N + 1}}\left( k \right),{\tilde R_{N - 1,N}}\left( k \right), \ldots ,{\tilde R_{12}}\left( k \right)$.

\item Find ${A_1}\left( k \right)$ from ${\tilde R_{12}}\left( k \right)$ by means of the top equation (\ref{eq:R1}).

\item Using ${A_1}\left( k \right)$ as an initial value, successively determine ${A_2}\left( k \right),{A_3}\left( k \right), \ldots ,{A_N}\left( k \right)$ from (\ref{eq:R7}).

\item For $n = 2,3, \ldots ,N$, use the known values of ${A_n}\left( k \right)$ and ${\tilde R_{n,n + 1}}\left( k \right)$ to determine ${B_n}\left( k \right)$ from the bottom equation (\ref{eq:R1}).

\item As long as $N \ne 1$, find ${B_{N + 1}}\left( k \right)$ from the equation ${B_{N + 1}}\left( k \right) = {T_{N,N + 1}}{B_N}\left( k \right)$. In the case $N=1$, set $B_2(k)=T_{12}e^{-kd_q}$.

\item The remaining unknowns (i.e. $A_{N+1}(k), B_1(k)$) are zero from (\ref{eq:A0}) and (\ref{eq:B0}).

\item Eq. (\ref{eq:gensolaltk}) now gives the solution ${\overline V _n}(k,z)$ in the $k$-domain, while (\ref{eq:gensol}) or (\ref{eq:gensolalt}) is the space-domain solution ${V_n}(\rho ,z)$ (in the form of an HT-integral).
\end{enumerate}

In Appendix~\ref{sec:charge-inside}, we discuss the more general case where the charge $q$ is placed within an arbitrary layer. 

\section{\label{sec:sec_iv} Two Layers ($N=1$); Meaning of Reflection and Transmission Coefficients}

\begin{figure}
\includegraphics{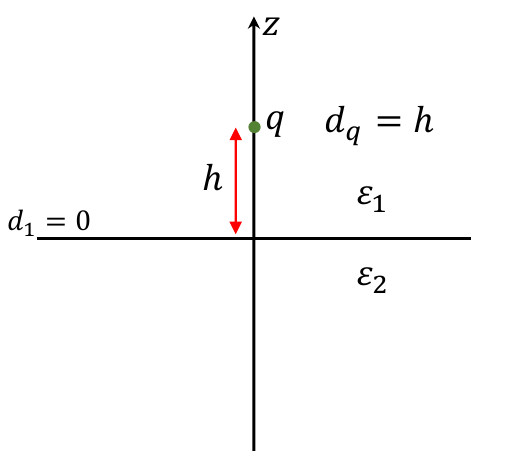}
\caption{\label{fig:fig_2} Two-layer problem ($N=1$), with origin placed at the boundary}
\end{figure}
We now turn to specific values of $N$, starting with the two-layer problem in which $N=1$. 
In Fig.~\ref{fig:fig_2}, we place the origin $z=0$ at the interface, and the charge $q$ at a distance $z=h$ above the origin, where $h>0$. Set $N=1$ (so that $n=1,2$), ${d_1} = 0$ and ${d_q} = h$ into the algorithm of Section~\ref{sec:sec_iii} to obtain ${B_1}(k) = {A_2}(k) = 0$, ${\tilde R_{12}}\left( k \right) = {R_{12}}$, ${A_1}(k) = {R_{12}}{e^{ - kh}},$ and ${B_2}(k) = {T_{12}}{e^{ - kh}}$. By~(\ref{eq:gensolaltk}), the $k$-domain solution is
\begin{eqnarray}
k{\kern 1pt} {\overline V _1}\left( {k,z} \right) = \frac{q}{{4\pi {\varepsilon _1}}}\left[ {{e^{ - k\left| {z - h} \right|}} + {R_{12}}{e^{ - k(h + z)}}} \right],\quad z > 0,
\label{eq:IV1}
 \\
k{\kern 1pt} {\overline V _2}\left( {k,z} \right) = \frac{q}{{4\pi {\varepsilon _2}}}{T_{12}}\,{e^{ - k(h - z)}} = \frac{q}{{4\pi {\varepsilon _1}}}{T_{21}}\,{e^{ - k(h - z)}},\quad z < 0.
\label{eq:IV2}
\end{eqnarray}
The last expression in (\ref{eq:IV2}) follows from the previous expression and (\ref{eq:coef2}). Eq. (\ref{eq:gensol}) provides the space-domain solution in integral form; and in this elementary problem, (\ref{eq:inv2}) can be used to evaluate the integrals. The result thus obtained is
\begin{eqnarray}
{V_1}(\rho ,z) = \frac{q}{{4\pi {\varepsilon _1}}}\left( {\frac{1}{{\sqrt {{\rho ^2} + {{(z - h)}^2}} }} + {R_{12}}\frac{1}{{\sqrt {{\rho ^2} + {{(z + h)}^2}} }}} \right),\quad z > 0
\label{eq:IV3},
 \\
{V_2}(\rho ,z) = \frac{q}{{4\pi {\varepsilon _2}}}{T_{12}}\frac{1}{{\sqrt {{\rho ^2} + {{(z - h)}^2}} }} = \frac{q}{{4\pi {\varepsilon _1}}}{T_{21}}\frac{1}{{\sqrt {{\rho ^2} + {{(z - h)}^2}} }},\quad z < 0
\label{eq:IV4}.
\end{eqnarray}
The well-known formulas (\ref{eq:IV3}) and (\ref{eq:IV4}) can be found in Jackson\cite{Jackson1} (for an HT-derivation, see Lebedev et. al.\cite{Lebedev1}). The first term in (\ref{eq:IV1}) (or (\ref{eq:IV3})) is the primal potential in the $k$-domain (or space domain). We will refer to the second term in (\ref{eq:IV1}) (or (\ref{eq:IV3})) as the reflected potential in the $k$-domain (or space domain). The solution in (\ref{eq:IV2}) (or (\ref{eq:IV4})) is the transmitted potential. By (\ref{eq:coef}), all three potentials (primal, reflected, transmitted) are finite when ${\varepsilon _1} \ne 0$ and ${\varepsilon _1}+{\varepsilon _2} \ne 0$.

Since there are no waves in our two-layer problem, there is no reflection or transmission in the usual (wave-propagation) sense. But we can use (\ref{eq:IV1})--(\ref{eq:IV4}) to physically interpret the quantities we called reflection and transmission coefficients: The reflected (transmitted) potential is generated by an image charge $qR_{12}$ (or $qT_{12}$) placed in the complementary region at $(\rho,z)=(0,-h)$ (or $(\rho,z)=(0,h)$) in an infinite medium with $\varepsilon=\varepsilon_1$ (or $\varepsilon=\varepsilon_2$). Furthermore, ${R_{12}}$ is the reflected-to-primal ratio at the boundary $z=0^+$, while ${T_{21}}$ is the transmitted-to-primal ratio at the boundary $z=0$; these two interpretations are true in both the space- and the $k$-domains.

\section{\label{sec:sec_v} Three Layers ($N=2$)}

\begin{figure}
\includegraphics{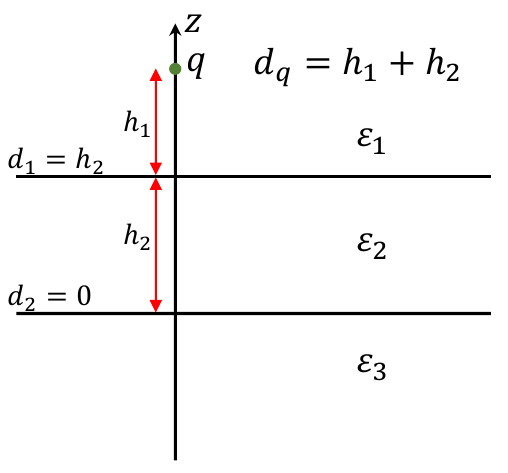}
\caption{\label{fig:fig_3} Three-layer problem ($N=2$), with origin placed at the 2-3 boundary}
\end{figure}

Our next problem is that of a three-layered region, as shown in Fig.~\ref{fig:fig_3}. Comparison to Fig.~\ref{fig:fig_1} and (\ref{eq:h}) gives $N=2$, ${d_2} = 0$, ${d_1} = {h_2}$, ${d_q} = {h_1} + {h_2}$. 

\subsection{Application of algorithm; the $\Psi$-function
}
The algorithm of Section \ref{sec:sec_iii} first gives the two generalized reflection coefficients as
\begin{equation}
{\tilde R_{23}}\left( k \right) = {R_{23}},\quad {\tilde R_{12}}\left( k \right) = \frac{{{R_{12}} + {R_{23}}\,{e^{ - 2k{h_2}}}}}{{1 + {R_{12}}{R_{23}}\,{e^{ - 2k{h_2}}}}}
\label{eq:v1},
\end{equation}
and then successively, ${A_1}\left( k \right) = {\textstyle{1 \over {D\left( k \right)}}}\left[ {{R_{12}}{e^{ - k({h_1} - {h_2})}} + {R_{23}}{e^{ - k({h_1} + {h_2})}}} \right]$, where the ``denominator'' $D(k)$ is 
\begin{equation}
D\left( k \right) = 1 - {R_{21}}{R_{23}}{e^{ - 2k{h_2}}};
\end{equation}
${A_2}\left( k \right) = \frac{1}{{D\left( k \right)}}{R_{23}}{T_{12}}{e^{ - k({h_1} + {h_2})}}$; ${B_2}\left( k \right) = \frac{1}{{D\left( k \right)}}{T_{12}}{e^{ - k({h_1} + {h_2})}}$; ${B_3}\left( k \right) = {\textstyle{1 \over {D\left( k \right)}}}{T_{12}}{T_{23}}{e^{ - k({h_1} + {h_2})}}$; and ${B_1}\left( k \right) = {A_3}\left( k \right) = 0$. Eq. (\ref{eq:gensol}) then yields the space-domain solutions in the three layers as
\begin{equation}
    V_1(\rho,z)=\frac{q}{{4\pi {\varepsilon _1}}}\Bigg[ {\frac{1}{{\sqrt {{\rho ^2} + {{\left( {z - {h_1} - {h_2}} \right)}^2}} }}+{R_{23}}\Psi\left( {\rho ,z + {h_1} + {h_2}} \right) + {R_{12}}\Psi\left( {\rho ,z + {h_1} - {h_2}} \right)} \Bigg],\  z > {h_2}
    \label{eq:v2}
\end{equation}

\begin{equation}
    {V_2}\left( {\rho ,z} \right) = \frac{q}{{4\pi {\varepsilon _2}}}T_{12}\left[ {{R_{23}}\Psi\left( {\rho ,z + {h_1} + {h_2}} \right) + \Psi\left( {\rho ,z - {h_1} - {h_2}} \right)} \right],
\quad 0 < z < {h_2}
\label{eq:v3}
\end{equation}
\begin{equation}
    {V_3}\left( {\rho ,z} \right) = \frac{q}{{4\pi {\varepsilon _3}}}{T_{12}}{T_{23}}\Psi\left( {\rho ,z - {h_1} - {h_2}} \right),{\kern 1pt} \quad z < 0
\label{eq:v4}
\end{equation}
In (\ref{eq:v2})--(\ref{eq:v4}) we introduced the \textit{normalized potential function} $\Psi\left( {\rho ,z - {z_0}} \right)$. This is an abbreviated notation for our function, the full notation being $\Psi\left( {\rho ,z - {z_0},2{h_2},{R_{21}}{R_{23}}} \right)$. In full notation, the general definition is
\begin{equation}
\Psi\left( {\rho ,z - {z_0},2h_2,R_{21}R_{23}} \right) = \int_0^\infty  {{J_0}(k\rho )\frac{{{e^{ - k\left| {z - {z_0}} \right|}}}}{{1 - R_{21}R_{23}{e^{ - 2kh_2}}}}dk} ,\quad 
\label{eq:v5}
\end{equation}
or equivalently,
\begin{equation}
{\Psi}\left( {\rho ,x,\Delta z,R} \right) = \int_0^\infty  {{J_0}(k\rho )\frac{{{e^{ - k\left| x \right|}}}}{{1 - R{e^{ - k\Delta z}}}}dk},\quad \Delta z>0, \quad x\in\mathbb{R},
\label{eq:v6}
\end{equation}
where we called $x = z - {z_0}$, $\Delta z=2h_2$, and $R=R_{21}R_{23}$.   Our notation (in particular, the meaning of $z_0$ and $\Delta z$) will become clearer in Section~\ref{sec:sec_viii}. For now, note that the two exponentials in (\ref{eq:v6}) (${e^{ - k\left| x \right|}}$ and ${e^{ - k\Delta z}}$) both decrease with $k$, except when $x=0$.

By (\ref{eq:inv2}) and (\ref{eq:v6}), $\Psi(R=0)$ is the normalized potential of a charge at $x=0$ (or $z = {z_0}$), 
\begin{equation}
    {\Psi}\left(\rho,x,\Delta z,{R = 0} \right) = \frac{1}{\sqrt {{\rho ^2} + x^2}}.
    \label{eq:asymp-1}
\end{equation}
This verifies that the solution (\ref{eq:v2})--(\ref{eq:v4}) reduces to that of the  two-layer problem (Section~\ref{sec:sec_vi}) when ${\varepsilon _3} = {\varepsilon _2}$ (so that $R_{23}=1-T_{23}=0$) or---with a proper renumbering---when ${\varepsilon_2} = {\varepsilon _1}$. 


\subsection{\label{sec:sec_vii} Validity of integral solutions; non-uniqueness when $R>1$}

We now discuss the convergence and interpretation of the integral  in (\ref{eq:v6}), or its equivalent (\ref{eq:v5}). Although we will soon consider complex $R_{21}R_{23}=R$, we initially assume $R \in \mathbb{R}$. When $R \in \mathbb{R}$, the integrand then has no pole\footnote{We ignore particular values of $\rho$ that correspond to Bessel-function zeros and render the integral convergent by cancelling the zero in the denominator.} on the integration path $\left[ {0, + \infty } \right)$  if and only if $R< 1$.
In this case, therefore, (\ref{eq:v6}) is a convergent integral and we have found a valid solution, which we believe to be unique. 

On the other hand, when the condition
\begin{equation}
    R=R_{21}R_{23}>1 \label{eq:Rgreaterthan1}
\end{equation}
is satisfied, there is a simple pole\footnote{It is readily checked that this pole is not cancelled whenever a sum of two $\Psi$-functions occurs, as in (\ref{eq:v2}).} at $k = {k_0}>0$, where $k_0$ is found from\footnote{Using the first (\ref{eq:coef}), we can show the equivalent to (\ref{eq:vii2}) equation 
\[
\varepsilon_3=-\varepsilon_2    \frac{\varepsilon_2 \tanh{kh_2}
+\varepsilon_1}{\varepsilon_2 +\varepsilon_1\tanh{kh_2}},
\]
which coincides with eqn. (3.300) of  Mayergoyz,\cite{Mayergoyz-book} where it is derived by other means.}
\begin{equation}
e^{2k_0h_2}=R_{21}R_{23}\quad \textrm{or}\quad {k_0} = \frac{{\ln (R_{21}R_{23})}}{{2h_2}}.  
\label{eq:vii2}
\end{equation}
Thus  no solution has been found in the case $R>1$. To deal with this case, we use a analytic-continuation argument which will demonstrate non-uniqueness and end up with a one-parameter family of solutions. Our argument, which is to some extent heuristic, will be strengthened and generalized in Section~\ref{sec:homogeneous}. 

\begin{figure}[t]
\includegraphics[width=11cm]{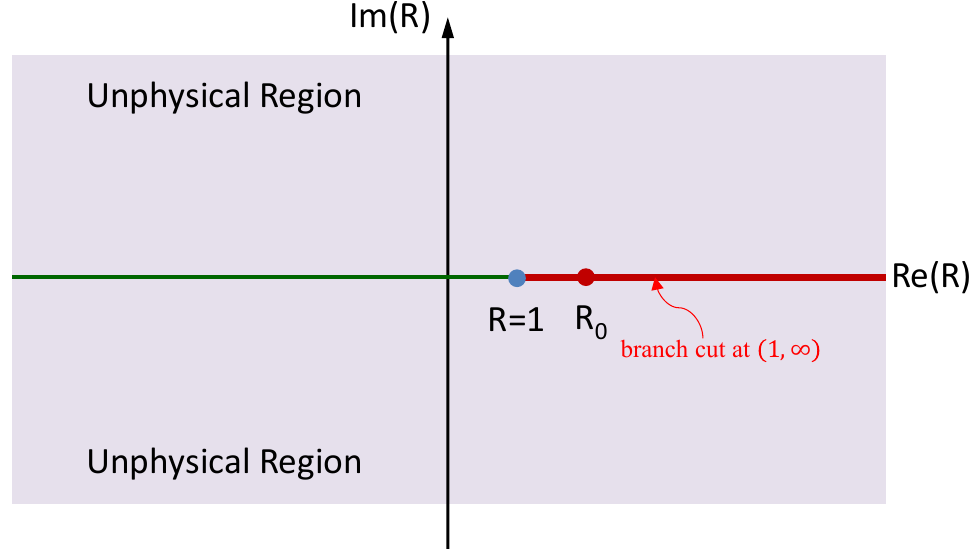}
\caption{\label{fig:fig_4} The solution $V^R$ in Section V.A holds for $R$ on the half line of the real axis $R<1$. In Section V.B, we  analytically continue this solution into the unphysical region $\mathbb{C\setminus R}$; and then, for any $R_0$ on the cut,  linearly combine the solutions $V^{R_0+i0}$ and $V^{R_0-i0}$ to obtain a one-parameter family of solutions to the boundary-value problem $\mathrm{BVP}^{R_0}$.}
\end{figure}

The first step (see Fig.~\ref{fig:fig_4}) is to remove the restriction $R \in \mathbb{R}$ and analytically continue the convergent-integral solution---which was originally valid subject to the condition $R<1$---into the complex  $R$-plane. This is possible as long as $R$ belongs to the cut plane according to 
\begin{equation}
R \in \mathbb{C} \setminus [1, + \infty )
\label{eq:vii3}.
\end{equation}

Subject to (\ref{eq:vii3}), the solution to the linear boundary-value problem defined by (\ref{eq:Poisson})--(\ref{eq:bcinfm}) is given by (\ref{eq:v2})--(\ref{eq:v4}), in which the various ${\Psi}$-functions  are still given by the convergent integral in (\ref{eq:v6}). Analytic continuation of the solution has thus amounted to analytic continuation of the $\Psi$-function, with the latter analytic continuation being apparent from (\ref{eq:v6}). 

Let us denote the said boundary-value problem by ${\rm{BV}}{{\rm{P}}^R}$ and its solution by ${V^R}$, so that ${V^R}$ is a $3 \times 1$ vector with components the ${V_1}(\rho ,z)$, ${V_2}(\rho ,z)$, ${V_3}(\rho ,z)$ of (\ref{eq:v2})--(\ref{eq:v4}). Given a complex value $R$,
${\rm{BV}}{{\rm{P}}^R}$ results by assigning suitable complex values to the three dielectric constants ${\varepsilon _1}$, ${\varepsilon _2}$, and ${\varepsilon _3}$.  Evidently, any desired $R \in \mathbb{C}$ can thus be obtained---but only in the special case $R\in\mathbb{R}$ and $\varepsilon _1, \varepsilon _2, \varepsilon _3 \in \mathbb{R}$  does ${\rm{BV}}{{\rm{P}}^R}$ correspond to an actual electrostatics problem. We thus use the term \textit{unphysical region} to designate the complex-$R$ plane with the real axis removed.

We now take an $R = {R_0}$ on the cut  (that is, an $R = {R_0}$  that is real with ${R_0}>1$), and consider ${\rm{BV}}{{\rm{P}}^{{R_0} + i\delta }}$ and ${\rm{BV}}{{\rm{P}}^{{R_0} - i\delta }}$, where $\delta  > 0$. As $\delta  \to 0$, (\ref{eq:Poisson})--(\ref{eq:bcinfm}) tell us that ${\rm{BV}}{{\rm{P}}^{{R_0} + i0}} = {\rm{BV}}{{\rm{P}}^{{R_0} - i0}} = {\rm{BV}}{{\rm{P}}^{{R_0}}}$, i.e., the two BVPs become identical. However the corresponding solutions are not the same (this non-uniqueness will soon be demonstrated explicitly) and will be denoted by ${V^{{R_0} + i0}}$ and ${V^{{R_0} - i0}}$; these are the limiting values of the solutions as we move in the unphysical region and arrive at the point $R_0$ of the cut from above and below, respectively. 

There are more solutions. Since our BVP is linear and inhomogeneous, any linear combination of ${V^{{R_0} + i0}}$ and ${V^{{R_0} - i0}}$ will also satisfy ${\rm{BV}}{{\rm{P}}^{{R_0}}}$, provided only that the respective complex coefficients sum to $1$. Any such linear combination can be written as
\begin{equation}
    \frac{V^{{R_0} + i0}+V^{{R_0} - i0}}{2}+
\alpha\,\frac{V^{{R_0} + i0}-V^{{R_0} - i0}}{2},\quad R_0>1,
    \label{eq:vii301}
\end{equation}
where $\alpha\in\mathbb{C}$. Evidently, the arbitrary constant $\alpha$ appears in the second term because this second term satisfies the \textit{homogeneous} ${\rm{BV}}{{\rm{P}}^{{R_0}}}$, which we denote by ${\rm{HBV}}{{\rm{P}}^{{R_0}}}$. ${\rm{HBV}}{{\rm{P}}^{{R_0}}}$ is specifically defined by (\ref{eq:Poisson})--(\ref{eq:bcinfm}), but with $q=0$. Throughout this paper, we only seek homogeneous solutions that $\varphi$-independent; that is, our ``homogeneous solutions'' are invariant with respect to rotations about the $z$-axis, on which $q$ lies.

In accordance with our previous analytic-continuation arguments, determination of the quantity in (\ref{eq:vii301}) amounts to finding the corresponding quantity involving the $\Psi$-function, viz., 
\begin{equation}
    \Psi_\textrm{NU}=\frac{\Psi({{R_0} + i0})+\Psi({{R_0} - i0})}{2}+
\alpha\,\frac{\Psi({{R_0} + i0})-\Psi({{R_0} - i0})}{2},\quad R_0\in (1,+\infty),
    \label{eq:vii302}
\end{equation}
where, for brevity, we suppress the first three arguments in ${\Psi }\left( {\rho ,x,\Delta z,R} \right)$ and where the subscript designates nonuniqueness. The two terms in (\ref{eq:vii302}) can be found from the integral expression (\ref{eq:v6}) by means of the two Plemelj formulas (see, for example, Ablowitz \& Fokas\cite{Fokas1}). Thus the first term (i.e. the mean) equals the $\Psi(R_0)$ given in (\ref{eq:v6}), but with the integral understood in the sense of the Cauchy principal value. We denote this quantity, which is real, by $\Psi_\textrm{PV}(R_0)$ or, in full notation, by $\Psi_\textrm{PV}(\rho ,x,\Delta z, R_0)$. Thus the first term in (\ref{eq:vii302}) equals
\begin{equation}
\Psi_\textrm{PV}\left( {\rho ,x,\Delta z,R} \right) = \dashint_0^\infty  {{J_0}(k\rho )\frac{{{e^{ - k\left| x \right|}}}}{{1 - R{e^{ - k\Delta z}}}}dk},\quad R\in (1,+\infty).
\label{eq:psipv}
\end{equation}
(We replaced the symbol $R_0$ by $R$). Appendix~\ref{sec:further-psi} demonstrates that the other Plemelj formula\cite{Fokas1} yields an nonzero expression for the quantity multiplying $\alpha$ in (\ref{eq:vii302}): Set $R_0=\exp(k_0\Delta z)$ in (\ref{eq:app-b2}) to obtain 
\begin{equation}
\Psi_\textrm{NU}=\Psi_\textrm{NU}( R,\beta)=  \Psi_\textrm{PV}(R)+\beta\,{J_0}\left( k_0\rho\right) e^{-k_0\abs{x}}, \quad R\in (1,+\infty),
\label{eq:vii4}
\end{equation}
where we replaced $R_0$ by $R$ and where the new constant $\beta$ is given by $\beta=i\pi\alpha/\Delta z$.

For $R\in\mathbb{R}$ with $R>1$, we have thus arrived at a one-parameter family of solutions to BVP$^R$. It is given by (\ref{eq:v2})--(\ref{eq:v4}), but with each $\Psi$ function replaced by the corresponding $\Psi_\textrm{NU}(R,\beta)$ of (\ref{eq:vii4}), in which $k_0$ is given  by (\ref{eq:vii2}). To have solutions that are real, we demand $\beta\in\mathbb{R}$. For brevity, we will sometimes omit the subscript $\textrm{PV}$ ($\Psi$ is to be understood as $\Psi_\textrm{PV}$ whenever $R>1$).

For $R>1$, an alternative to (\ref{eq:psipv}) expression that appears to be more suitable for numerical evaluation is found in Appendix~\ref{sec:further-psi}. It is 
\begin{equation}
{\Psi}_\mathrm{PV}\left( {\rho ,x,\Delta z,R} \right) = \frac{1}{\Delta z}\dashint_1^\infty  {J_0}\left(\frac{\rho}{\Delta z}\ln{t} \right)\frac{t^{-\frac{\abs{x}}{\Delta z}}}{t-R}\,dt,\quad R\in (1,+\infty).
\label{eq:psipv-alternative}
\end{equation}

\subsection{The homogeneous solution explicitly}

The above prescription for obtaining the non-unique solution implies a similar one for the ($\varphi$-independent) homogeneous solution, which we now denote by $V_H(\rho,z)$ (i. e. $V_H(\rho,z)$ is a $3 \times 1$ vector with components ${V_{1H}}(\rho ,z)$, ${V_{2H}}(\rho ,z)$, and ${V_{3H}}(\rho ,z)$): To obtain $V_H(\rho,z)$, ignore the first term in (\ref{eq:v2}); and in (\ref{eq:v2})--(\ref{eq:v4}), replace each $\Psi(\rho,x)$ by $J_0(k_0\rho) e^{-k_0|x|}$. It follows that ${V_{1H}}(\rho ,z)$ is a constant times $J_0(k_0\rho)e^{-k_0z}$, and it is convenient to set 
\begin{equation}
    V_{1H}(\rho,z)=V_HJ_0(k_0\rho)e^{-k_0z},\quad z>h_2,
    \label{eq:vii401}
\end{equation}
where $V_H$ is an arbitrary real constant. With this normalization, our prescription yields\footnote{The expression for $V_H$ is $V_H=qe^{-k_0(h_1+h_2)}(R_{23}+R_{12}e^{2k_0h_2})/(4\pi\varepsilon_1)$ (the factor $q$ being insignificant). Note that this expression---as well as (\ref{eq:vii402}) and (\ref{eq:vii403})---can be slightly simplified via the second equation (\ref{eq:vii2}).}
\begin{equation}
    V_{2H}(\rho,z)=V_HT_{21}J_0(k_0\rho)\frac{e^{k_0z}+R_{23}e^{-k_0z}}{R_{23}+R_{12}e^{2k_0h_2}},\quad 0<z<h_2,
    \label{eq:vii402}
\end{equation}
\begin{equation}
    V_{3H}(\rho,z)=V_HT_{21}T_{32}J_0(k_0\rho)\frac{e^{k_0z}}{R_{23}+R_{12}e^{2k_0h_2}},\quad z<0.
    \label{eq:vii403}
\end{equation}
These equations, which involve no integrals, can be verified directly. In other words, with the $k_0$ given in (\ref{eq:vii2}), we can easily see that (\ref{eq:vii401})--(\ref{eq:vii403}) satisfy (\ref{eq:Poisson})--(\ref{eq:bcinfm}) for the special case $q=0$. 

When $R=R_{21}R_{23}=1$, the pole at $k = {k_0}$ moves to $k = 0$, which is the endpoint of integration in (\ref{eq:v5}). In this critical case (\ref{eq:v2})--(\ref{eq:v4}) have no meaning as a solution. We will also consider the points $R =  \pm \infty $ as being critical.

\section{\label{sec:sec_vi} Two Layers above a Ground}

\begin{figure}
\includegraphics{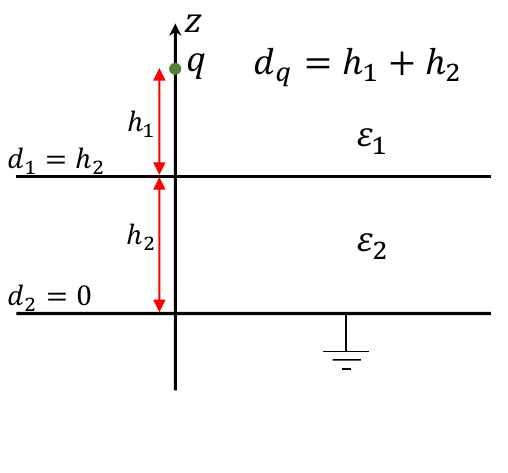}
\caption{\label{fig:fig_5} Two layers above a ground plane}
\end{figure}
We now turn to the problem of Fig.~\ref{fig:fig_5}, which is a special case that illustrates most of the arguments of Section~\ref{sec:sec_v} with fewer algebraic complications. Fig.~\ref{fig:fig_5} is the configuration of Fig.~\ref{fig:fig_3} in the limit ${\varepsilon _3} \to \infty $, so that $R_{23}=-1=1-T_{23}$ by (\ref{eq:coef}), and ${V_3}\left( {\rho ,z} \right) = 0$ by (\ref{eq:v4}), as expected.  The solutions in layers 1 and 2 are then given by the limiting values of the solutions in (\ref{eq:v2}) and (\ref{eq:v3}). They are
\begin{equation}
    V_1\left( {\rho ,z} \right) = \frac{q}{{4\pi {\varepsilon _1}}}\left[ {\frac{1}{{\sqrt {{\rho ^2} + {{\left( {z - {h_1} - {h_2}} \right)}^2}} }} - {\Psi}\left( {\rho ,z + {h_1} + {h_2}} \right) + {R_{12}}{\Psi }\left( {\rho ,z + {h_1} - {h_2}} \right)} \right],\quad z > {h_2}
\label{eq:vi1}
\end{equation}
\begin{equation}
    {V_2}\left( {\rho ,z} \right) = \frac{q}{{4\pi {\varepsilon _2}}}{T_{12}}\Bigg[ {{\Psi}\left( {\rho ,z - {h_1} - {h_2}} \right) - {\Psi}\left( {\rho ,z + {h_1} + {h_2}} \right)} \Bigg],\quad 0 < z < {h_2}
\label{eq:vi2}
\end{equation}
where, now, it is to be understood that the four ${\Psi}\left( {\rho ,z - {z_0}} \right)$  are abbreviated notations for ${\Psi}\left( {\rho ,z - {z_0},2{h_2},{R_{12}}} \right)$, i.e., set $\Delta z=2h_2$ and $R=R_{12}$ in (\ref{eq:v6}). All integrals are convergent, and the above solution is valid, if and only if $R=R_{12}<1$. These solutions are also given, in different algebraic form, in the work of Lebedev et. al.\cite{Lebedev1}

The integrals diverge, and (\ref{eq:vi1})--(\ref{eq:vi2}) are invalid, when $R=R_{12}>1$. Our analytic continuation arguments (of Section~\ref{sec:sec_v}) carry over verbatim and, once again, lead to a nonunique solution.  Thus when $R_{12}>1$, there is a simple pole at $k_0=\ln(R_{12})/(2h_2)$ (this equation amounts to $\tanh(k_0h_2)=-\varepsilon_2/\varepsilon_1)$, and we end up with a one-parameter family of solutions by replacing the various $\Psi$ functions in (\ref{eq:vi1})--(\ref{eq:vi2}) by the corresponding $\Psi_\textrm{NU}(R_{12},\beta)$ given in
(\ref{eq:vii4}). In the nonunique solution resulting from this procedure, the homogeneous part $V_H(\rho,z)$ consists of all terms multiplied by the factor $\beta$. Normalizing as in Section~\ref{sec:sec_v}, we find this $V_H(\rho,z)$ as
\begin{equation}
V_{1H}(\rho,z)=V_HJ_0(k_0\rho)e^{-k_0z},\quad z>h_2,
\label{eq:vi3}
\end{equation}
\begin{equation}
V_{2H}(\rho,z)=V_He^{-k_0h_2}J_0(k_0\rho)\frac{\sinh(k_0z)}{\sinh(k_0h_2)},\quad 0<z<h_2.
\label{eq:vi4}
\end{equation}
where $V_H$ is an arbitrary real constant. This $V_H(\rho,z)$ is immediately verified directly: Continuity at $z=h_2$ is obvious, while continuity of $\varepsilon_n\partial V_{nH}$ amounts to the above-noted equation $\tanh(k_0h_2)=-\varepsilon_2/\varepsilon_1$. Note, finally, that (\ref{eq:vi3}) and (\ref{eq:vi4}) also follow by specializing (\ref{eq:vii2}), (\ref{eq:vii401}), and (\ref{eq:vii402}).

\section{On-axis values and Lerch's transcedent}

The discussions in this and the next two sections concern the ${\Psi}$-function and are applicable to both specific problems we discussed (Sections~\ref{sec:sec_v}~and~\ref{sec:sec_vi}).

Lerch's transcedent, also called the Hurwitz-Lerch zeta-function $\Phi(R,n,y)$, is defined\footnote{Eqn. (\ref{eq:phidef}) actually specializes a more general definition\cite{NIST, Wolfram} which allows $y,n\in\mathbb{C}$.} for $|R|<1$ by the series\cite{NIST, Wolfram}
\begin{equation}
\label{eq:phidef}
    \Phi(R,n,y)=\sum_{m=0}^\infty\frac{R^m}{(m+y)^n}, \quad |R|<1,\quad y>0,\quad n=1,2,\ldots,
\end{equation}
and for other values of $R\in\mathbb{C}$ by analytic continuation. $\Phi$ has the integral representation\cite{NIST, Wolfram}
\begin{equation}
\label{eq:phiintegralrepresentation}
    \Phi(R,n,y)=\frac{1}{(n-1)!}\int_0^\infty\frac{u^{n-1}e^{-yu}}{1-Re^{-u}}\,du, \quad R \in \mathbb{C} \setminus [1, + \infty ).
\end{equation}
For the case $R>1$, we use the symbol $\Phi_\mathrm{PV}(R,n,y)$ to denote the above integral understood in the sense of the Cauchy principal value, so that 
\begin{equation}
\label{eq:phiintegralrepresentation-pv}
    \Phi_\mathrm{PV}(R,n,y)=\frac{1}{(n-1)!}\,\dashint_0^\infty\frac{u^{n-1}e^{-yu}}{1-Re^{-u}}\,du, \quad R \in (1, + \infty ).
\end{equation}
Thus by the first Plemelj formula\footnote{An alternative to (\ref{eq:phiplemelj}) expression is $\Phi_\mathrm{PV}(R,n,y)=\Re{\Phi(R,n,y)}$, where the $R$ in the right-hand side stands for either $R+i0$ or $R-i0$. The choice is immaterial because $\Phi(R+i0,n,y)$ and $\Phi(R-i0,n,y)$ are complex conjugates.}
\begin{equation}
\label{eq:phiplemelj}
    \Phi_\mathrm{PV}(R,n,y)=\frac{\Phi(R+i0,n,y)+\Phi(R-i0,n,y)}{2}, \quad R \in (1, + \infty ).
\end{equation}

We show in Appendix~\ref{sec:further-psi} that $ \Psi\Delta z$ can be expanded into powers of $\frac{\rho}{\Delta z}$ according to
\begin{equation}
\label{eq:psi7}
    \Psi(\rho,x,\Delta z,R)=\frac{1}{\Delta z}\sum_{m=0}^\infty \left(-\frac{1}{4}\right)^m\binom{2m}{m}\,\Phi\left(R,2m+1,\frac{|x|}{\Delta z}\right)\left(\frac{
    \rho}{\Delta z}\right)^{2m},
\end{equation}
where $\binom{2m}{m}=\frac{(2m)!}{(m!)^2}$ is the binomial coefficient. By (\ref{eq:psi7}), the on-axis value of $\Psi$ is
\begin{equation}
\label{eq:psionaxis}
    \Psi(\rho=0,x,\Delta z,R)=\frac{1}{\Delta z} \Phi\left(R,1,\frac{|x|}{\Delta z}\right).
\end{equation}
Eqn. (\ref{eq:psionaxis}) also follows by comparing (\ref{eq:phiintegralrepresentation}) to (\ref{eq:app-b0}). In (\ref{eq:psi7}) and (\ref{eq:psionaxis}), $\Psi$ and $\Phi$ are to be understood as $\Psi_\mathrm{PV}$ and $\Phi_\mathrm{PV}$ whenever $R>1$.

To summarize, our normalized potential-function~$\Delta z\Psi(\rho,x,\Delta z,R)$ is a generalization of its on-$z$-axis value $\Phi\left(R,1,\frac{|x|}{\Delta z}\right)$ which in turn is (a special case of) a well-studied mathematical function with many known properties.\cite{NIST,Wolfram} Furthermore, the off-$z$-axis values of $\Delta z\Psi$ are given by the series  (\ref{eq:psi7}), which involves $\Phi$ functions of the more general form (\ref{eq:phidef}).

\section{\label{sec:sec_viii} Image-series solution; $R$-regions}

We now discusss ${\Psi}\left( {\rho ,z - {z_0},\Delta z,R} \right)$, where $(\rho,z)$ is the observation point. Subject to the condition $ - 1 \le R < 1$, we can apply the geometric series and (\ref{eq:inv2}) to the integral in (\ref{eq:v6}) to obtain
\begin{equation}
{\Psi}\left( {\rho ,z - {z_0},\Delta z,R} \right) = \sum\limits_{m = 0}^\infty  {\frac{R^m}{{\sqrt {{\rho ^2} + {{\left(z - z_m\right)}^2}} }}} ,\quad  - 1 \le R < 1,
\label{eq:viii1}
\end{equation}
where the symbol $z_m$ is defined by
\begin{equation}
\label{zm-definition}
    z_m=
    \begin{cases}
        z_0-m\Delta z,\quad z\ge z_0,\\
        z_0+m\Delta z, \quad z\le z_0, 
    \end{cases}
    \quad m\in\mathbb{Z},
\end{equation}
so that ${\left( {\left| {z - {z_0}} \right| + m\Delta z} \right)}^2=(z-z_m)^2$. Note from (\ref{eq:psionaxis}) that (\ref{eq:viii1}) is a generalization of the well-known series (\ref{eq:phidef}) for the case $n=1$. 

\begin{figure}
\includegraphics{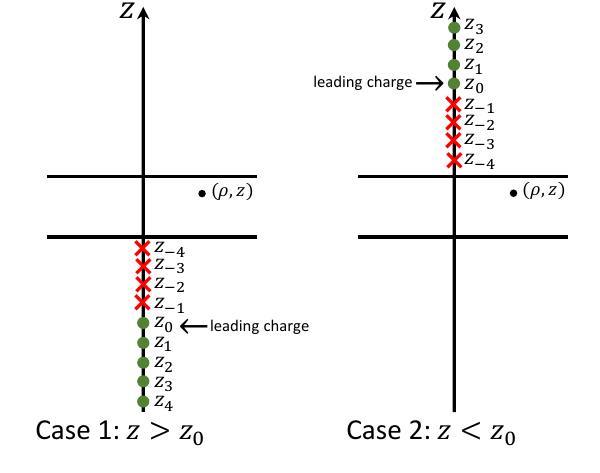}
\caption{\label{fig:fig_6} Image charges (marked with dots) and phantom charges (marked with Xs) for the cases $z>z_0$ (left) and $z<z_0$ (right). Note that there is an infinite number of image charges (at ${z_0},{z_1},{z_2}, \cdots $) but only a finite number of phantom charges (at ${z_{ - 1}},{z_{ - 2}}, \cdots ,{z_{ - M}}$). The figure takes $M=4$.}
\end{figure}

Eqns. (\ref{eq:viii1}) and (\ref{zm-definition}) tell us that the source of our normalized  potential $\Psi\left( {\rho ,z - {z_0},\Delta z,R} \right)$ can be viewed a one-dimensional, semi-infinite lattice of normalized equispaced image charges, shown as large dots ($\bullet$) in Fig.~\ref{fig:fig_6}. The \textit{lattice spacing} (distance between adjacent charges) is $\Delta z$. Charge number $m$ corresponds to the $m$th term of the sum, is located at $z = z_m$, and its normalized charge is ${R^m}$. The \textit{leading charge} is located at $z = z_0$ and corresponds to $m=0$. When $z>z_0$ the lattice extends downwards (to $z=-\infty$) and the observation point $\left( {\rho ,z} \right)$ is located above the lattice; but when  $z<z_0$, the lattice extends upwards (to $z=+\infty$) and the observation point is located below it. Of course the lattice location is the same (above or below $\left( {\rho ,z} \right)$)  for all $\left( {\rho ,z} \right)$ within a particular layer, as $z-z_0$ does not change sign. In the first term of (\ref{eq:vi2}), for example, $z-h_1-h_2<0$ for all $z$ with $0<z<h_2$, corresponding to a lattice that lies entirely above the layer (as in the right Fig.~\ref{fig:fig_6}). We refer to the series in (\ref{eq:viii1}) as the \textit{image series}.

These results are relevant only when $ - 1 \le R < 1$; otherwise, the series in (\ref{eq:viii1}) diverges. We have already called $\mathbb{C}\setminus\mathbb{R}$ the unphysical region (of the complex-$R$ plane). Depending on the type of solution and the form of the pertinent $\Psi$ function, we can further divide the real-$R$ axis in the following manner.

\begin{enumerate}

\item CI-CS region (convergent integral-convergent series region), $ - 1 \le R < 1$: The involved $\Psi$-functions can be calculated from the defining integral (\ref{eq:v6}) or, alternatively,  from the image series (\ref{eq:viii1}); both these expressions converge.  

\item CI-DS region (convergent integral-divergent series region), $R<-1$: Our solution can be calculated via (\ref{eq:v6}), as the integral converges. The image series (\ref{eq:viii1}), however, diverges.

\item NU-DS region, $R>1$: This is the non-unique region, where there is a one-parameter family of solutions. The integral (\ref{eq:v6}) defining $\Psi$ diverges (unless interpreted in the principal-value sense), and so does the image series in (\ref{eq:viii1}). Here, we must replace all involved $\Psi$-functions (e.g. in (\ref{eq:vi1})--(\ref{eq:vi2})) by the corresponding $\Psi_\textrm{NU}(R,\beta)$ given in (\ref{eq:vii4}), in which the Cauchy principal value integral (\ref{eq:psipv-alternative}) appears, corresponding to our particular solution. 
\item Critical points, $R = 1$ or $R =  \pm \infty $: No solution has been found.
\end{enumerate}

In the CI-DS regions, we can regard the integral expression (\ref{eq:v6}) as the regularization of  the divergent series (\ref{eq:viii1}) (namely of the image series, which only converges in the CI-CS region).

\begin{figure}
\includegraphics{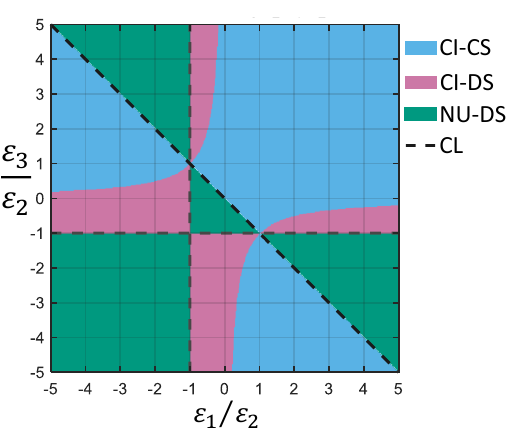}
\caption{\label{fig:fig_7} Regions for three-layer  problem ($N=2$) of Section V.}
\end{figure}

For the problems of Sections~\ref{sec:sec_v} and \ref{sec:sec_vi}, we have $R = {R_{21}}{R_{23}}$ and  $R = {R_{12}}$, respectively. We can thus use the first equation (\ref{eq:coef}) and describe the various regions in terms of ratios of dielectric constants. In the simpler problem of Section~\ref{sec:sec_vi}, the NU-DS, CI-DS, and CI-CS regions are $\frac{{{\varepsilon _1}}}{{{\varepsilon _2}}} <  - 1$,  $ - 1 < \frac{{{\varepsilon _1}}}{{{\varepsilon _2}}} < 0$, and  $0 \le \frac{{{\varepsilon _1}}}{{{\varepsilon _2}}}$, respectively, with the critical points at $\frac{{{\varepsilon _1}}}{{{\varepsilon _2}}} =  \pm \infty $ and $\frac{{{\varepsilon _1}}}{{{\varepsilon _2}}} =  - 1$. In the problem of Section~\ref{sec:sec_v}, we can depict the regions in a plane with a horizontal axis $\frac{{{\varepsilon _1}}}{{{\varepsilon _2}}}$ and a vertical axis $\frac{{{\varepsilon _3}}}{{{\varepsilon _2}}}$, see Fig.~\ref{fig:fig_7}. In this case, it is more appropriate to speak of critical lines (CL) instead of critical points. The figure is a bit simpler for practical situations, because layers that extend to infinity usually have positive dielectric constants. 

\section{\label{sec:phantom} Phantom images}

The image series in (\ref{eq:viii1}) was to be expected. Indeed, the corresponding solutions to the problems of Sections~\ref{sec:sec_v} and \ref{sec:sec_vi} can be derived from first principles, taking a cue from the simple, one-boundary problem of Section~\ref{sec:sec_iv}: In the case of two boundaries, we take the image of $q$ with respect to the two, then the images of the images, and so on. One thus obtains a solution involving several infinite image series, but then finds that the obtained series converge only subject to the condition $-1\le R<1$ (CI-CS region). Only then is this elementary image theory applicable.

Here, we turn to the case $|R|>1$ and develop  a different---and less intuitive---sort of image theory, applicable to the solution when $R<-1$ (CI-DS region), and to our  particular solution when $R>1$ (NU-DS region). The obtained formulas are asymptotic rather than exact, as they are valid when $\abs{R}$ is large. Furthermore our theory does not always hold; but when it does, the image locations are different than before.

\subsection{Finte sum of phantom images}

As $R\to -\infty$ (in the CI-DS region), or as $R\to +\infty$ (in the NU-DS region), we can apply the asymptotic formula (\ref{eq:asymp-2}) to the $\Psi$-function on the right-hand side of (\ref{eq:app-b4}) to obtain
\begin{equation}
{\Psi}\left( {\rho ,z - {z_0},\Delta z,R} \right) = \sum\limits_{m = -M}^{-1}  \frac{{{-R^{m}}}}{{{\sqrt {{\rho ^2} + {{\left( z-z_m \right)}^2}} }}}+O\left(\frac{1}{R^{M+1}}\right),\textrm{\ as \ }R\to\pm\infty,
\label{eq:phantom}
\end{equation}
 where $z_m$ is given by (\ref{zm-definition}). By the condition in (\ref{eq:app-b4}), the asymptotic formula (\ref{eq:phantom}) is valid when
 $\abs{z-z_0}-(M+1)\Delta z>0$, a condition amounting to 
\begin{equation}
    \big(z>z_{-M-1}\  \mathrm{if}\  z>z_0\big)\quad \mathrm{or}\quad
    \big(z<z_{-M-1}\  \mathrm{if} \  z<z_0\big),\quad M=0,1,2,\ldots.
    \label{eq:phantom-condition}
\end{equation}
Assume that the sum in (\ref{eq:phantom}) is non-empty, so that the $O\left(\frac{1}{R^{M+1}}\right)$ is meaningful as a remainder (when we approximate $\Psi$ by the sum). Then the sum is best understood via a detailed comparison to the image series of (\ref{eq:viii1}), as follows.
\begin{itemize}
    \item The $R$-regions of validity are different: Eq. (\ref{eq:viii1}) holds (exactly) in the CI-CS region, whereas (\ref{eq:phantom}) holds (for large $\abs{R}$) in the CI-DS region (for $\Psi$) and, also, in the NU-DS region (for $\Psi_\textrm{PV}$, corresponding to the particular solution).
    \item The image series of Section~\ref{sec:sec_viii} holds for all legitimate values of $\rho$, $z-z_0$, and $\Delta z$, whereas (\ref{eq:phantom}) holds when (\ref{eq:phantom-condition}) is satisfied, as further discussed below.
    \item There is an infinite number of sources in (\ref{eq:viii1}), but $M$ sources in (\ref{eq:phantom}). Thus the image series of the CI-CS region is associated with a semi-infinite lattice, whereas the lattice corresponding to (\ref{eq:phantom}) is finite.  The lattice spacing is $\Delta z$ in both cases.
    \item To obtain the finite lattice of (\ref{eq:phantom}), we \textit{extend} the semi-infinite one of (\ref{eq:viii1}), as shown in Fig.~\ref{fig:fig_6} (which assumes $M=4$). We extend upward (from the leading charge at $z=z_0$) if $z>z_0$, and downward if $z<z_0$.  Since the source locations corresponding to (\ref{eq:phantom}) differ (from the locations of the sources we usually regard as ``images''), we call the sources/images in (\ref{eq:phantom}) ``phantom.'' They are symbolized by Xs ($\times$) in Fig.~\ref{fig:fig_6}.   
    \item Suppose we move away from the leading charge at $z=z_0$. Then the normalized image charges (corresponding to (\ref{eq:viii1})) are $R,R^2, R^3,\ldots$, while in (\ref{eq:phantom}), the normalized phantom image charges are $-R^{-1},-R^{-2},-R^{-3},\ldots, -R^{-M}$. In the latter case, the dominant term in the sum corresponds to $m=-1$, i.e. to the phantom image  that is \textit{farthest} from the layer. The condition (\ref{eq:phantom-condition}) means that there is room for an additional charge (at $z=z_{-M-1}$) between the last phantom charge and $z$.
    \item Let us for the moment ignore the condition (\ref{eq:phantom-condition}). Then (\ref{eq:phantom}) has the usual form of an asymptotic power series, i.e., the remainder is of the order of the first neglected term. Furthermore, if we set $M=+\infty$ in the finite sum in (\ref{eq:phantom}), we obtain the infinite series $S(\rho,z)$ given by
\begin{equation}
 S(\rho,z)=\sum\limits_{m = -\infty}^{-1}  \frac{{{-R^{m}}}}{{{\sqrt {{\rho ^2} + {{\left( z-z_m \right)}^2}} }}}.
\end{equation}
For $|R|>1$, it happens that $S(\rho,z)$ is \textit{convergent}. It is thus natural to inquire whether ${\Psi}\left( {\rho ,z - {z_0},\Delta z,R} \right)=S(\rho,z)$ in the sense of a true equality, i.e., if (\ref{eq:phantom}) is  not merely an asymptotic relation. The answer is no, because $S(0,z)\to\infty$ as $z\to z_m$, whereas $\Psi(0,z_m-z_0,\Delta z,R)$ is certainly finite. Physically, the series $S$ amounts to extending the finite phantom-image lattice beyond the observation point; as this point can now approach a source, equality should not be expected. In the on-axis case $\rho=0$, our assertions about $S(0,z)$ can be rephrased as statements about Lerch's transcedent $\Phi(R,1,\frac{|x|}{\Delta z})$, see (\ref{eq:psionaxis}). Indeed  (a generalization of) the series $S(0,z)$ appears in a recent study\cite{Daalhuis} of the asymptotics of $\Phi$; see also eqn. 25.14.9 of Ref.~\onlinecite{NIST}. 
   \end{itemize}

\subsection{Choice of $M$; numerical accuracy}

If no $M$ satisfies (\ref{eq:phantom-condition}), then (\ref{eq:phantom}) does not hold. Otherwise, let 
$M_\mathrm{max}$ be the largest $M$ satisfying (\ref{eq:phantom-condition}), so that the choice $M=M_\mathrm{max}$ in (\ref{eq:phantom}) is optimal. If $M_\mathrm{max}=0$, (\ref{eq:phantom}) reduces to the asymptotic relation (\ref{eq:asymp-2}) and yields little essential information.

For a given observation point $(\rho,z)$, having a meaningful phantom-image sum (i.e, a non-empty sum in (\ref{eq:phantom})) is tantamount to $M_\mathrm{max}\ge 1$. The accuracy can be very good even when 
$M_\mathrm{max}$ is small. To illustrate, when $\rho=2.3$, $z=3.2$, $z_0=0.3$, and $\Delta z=0.7$, then $M_\mathrm{max}=3$; in this case, a three-term phantom-image sum ($M=M_\mathrm{max}=3$) yields a $1.1\%$ error (compared to the numerical evaluation of $\Psi$) when $R=-5$, and a $0.14\%$ error when $R=-10$. For positive $R$, the corresponding errors for $R=5$ and $R=10$ are $1.3\%$ and $0.3\%$, respectively. If we decrease $z$ to the value $1.9$, we have $M_\mathrm{max}=1$ and a $7\%$ error when $R=20$. In this case, adding more terms does not help; but our one-term approximation suffices for a back-of-the-envelope calculation.

To have an approximation that remains \textit{continuous} within a given layer (say for all $z$ with $0<z<h_2$ in Fig.~\ref{fig:fig_5}), $M$ should be such that the full lattice of phantom images---plus one more---remains entirely below (or above) the layer.

\section{\label{sec:homogeneous} Homogeneous solution for more layers}

Subject to the condition $R_{21}R_{23}>1$, our analytic-continuation arguments of Section~\ref{sec:sec_vii} led to an explicit, $\varphi$-independent solution to the \textit{homogeneous} three-layer problem ($N=2$).\footnote{When $\varepsilon_1+\varepsilon_2=0$, even the two-layer problem ($N=1$, Section~\ref{sec:sec_iv}) has a $\varphi-$independent homogeneous solution. Perhaps unexpectedly, it is quite general and immediately verifiable from (\ref{eq:Poisson})--(\ref{eq:bcinfm}). It is $V_{1,H}(\rho,z)=f(\rho,z)$ and $V_{2,H}(\rho,z)=f(\rho,-z)$, where $f(\rho,z)$ is any solution to the $\varphi$-independent Laplace equation that also satisfies $f(\rho,+\infty)=0$. Viewed otherwise, 
$V_{1,H}(\rho,z)=V_HJ_0(k\rho)e^{-kz}$ and $V_2(\rho,z)=V_HJ_0(k\rho)e^{kz}$ satisfy the homogeneous problem for any $k>0$ and $V_H\in\mathbb{R}$, and so do their superpositions over $k$. But, by (\ref{eq:gen}), such superpositions can represent \textit{any} of the above-defined $f(\rho,z)$ and $f(\rho,-z)$.} With this solution known, it is now easy to find a similar one for $N$ layers ($N\ge 2$), i.e., a solution to the general problem of Fig.~\ref{fig:fig_1} for the case where $q=0$. No analytic-continuation argument is needed, as we can modify the rudimentary discussions in Section \ref{sec:sec_iii}. We simply postulate a solution to the homogeneous problem of the form 
\begin{equation}
    V_{1,H}(\rho,z)=V_HJ_0(k\rho)e^{-kz}, \quad z>d_1,
\end{equation}
\begin{equation}
    V_{n,H}(\rho,z)=V_HJ_0(k\rho)\left(\alpha_ne^{-kz}+\beta_ne^{kz}\right), \quad d_n<z<d_{n-1}\quad (n=2,3,\ldots,N),
\end{equation}
\begin{equation}
    V_{N+1,H}(\rho,z)=V_HJ_0(k\rho)\beta_{N+1}e^{kz},\quad z<d_N,
\end{equation}
where $V_H$ is an arbitrary real constant, $\alpha_n$ and $\beta_n$ are real constants to be determined, and the ``eigenvalue'' $k$ is a positive constant to be determined. Our postulated solution already satisfies the (homogeneous) Laplace equation in each layer and (since $k>0$) the conditions (\ref{eq:bcinfp}) and (\ref{eq:bcinfm}) at $z=\pm\infty$, so it suffices to enforce the boundary conditions (\ref{eq:bc1}) and (\ref{eq:bc2}) at $z=d_1, d_2,\ldots,d_N$. Evidently, the common factor $V_HJ_0(k\rho)$ plays no role in this procedure. 

The two boundary conditions at $z=d_1$ give the two equations
\begin{equation}
\label{eq:hom1}
    \alpha_2+\beta_2e^{2kd_1}=1,\quad \varepsilon_2\alpha_2-\varepsilon_2\beta_2e^{2kd_1}=\varepsilon_1,
\end{equation}
which can be solved for $\alpha_2$ and $\beta_2$ if we assume (initially) that $k$ is known. Similarly, the two boundary conditions at $z=d_2,\ldots,d_{N-1}$ give 
\begin{equation}
\label{eq:hom2} 
\begin{split}
\alpha_{n+1}+\beta_{n+1}e^{2kd_n}&=\alpha_n+\beta_ne^{2kd_n},\\ 
\varepsilon_{n+1}\alpha_{n+1}-\varepsilon_{n+1} \beta_{n+1}e^{2kd_n}&=\varepsilon_n\alpha_n-\varepsilon_n \beta_ne^{2kd_n},
\end{split}
\quad n=2,\ldots,N-1.
\end{equation}
For each $n$, we regard (\ref{eq:hom2}) as a $2\times 2$ system for $\alpha_{n+1}$ and $\beta_{n+1}$. Since $\alpha_2$, $\beta_2$ are already known (in terms of $k$), we can successively find $\alpha_3$, $\beta_3$; $\alpha_4$, $\beta_4$;$\ldots$; and, lastly, $\alpha_N$, $\beta_N$.
The remaining two boundary conditions (at $z=d_N$) yield
\begin{equation}
\label{eq:hom3}
\alpha_N+\beta_Ne^{2kd_N}=\beta_{N+1}e^{2kd_N},\quad \varepsilon_N\alpha_N-\varepsilon_N\beta_Ne^{2kd_N}=-\varepsilon_{N+1}\beta_{N+1}e^{2kd_N},
\end{equation}
from which we can eliminate $\beta_{N+1}$ to obtain
\begin{equation}
\label{eq:hom4}  \alpha_N(\varepsilon_N+\varepsilon_{N+1})=\beta_Ne^{2kd_N}(\varepsilon_N-\varepsilon_{N+1}).
\end{equation}
The next step is to substitute the known (in terms of $k$) values $\alpha_N$ and $\beta_N$ into (\ref{eq:hom4}), thus obtaining an equation for $k$, henceforth termed ``eigenvalue equation.'' For each positive solution (i.e., for each eigenvalue $k$) we can finally determine $\alpha_n$ and $\beta_n$ as described above.

The eigenvalue equation itself is more easily found by dealing with the ratios $\alpha_n/\beta_n$ or, more conveniently, with $\tilde{r}_n\overset{\Delta}{=}\frac{\alpha_n\exp(-2kd_n)}{\beta_n}$ ($n=2,3,\ldots,N$) (compare to the Stokes-like generalized reflection coefficients in the bottom equation (\ref{eq:R1})): Eqns. (\ref{eq:hom1}) and (\ref{eq:hom4}) give
\begin{equation}
\label{eq:hom5}
\tilde{r}_2=\frac{-1}{R_{1,2}}\,e^{2kh_2},
\end{equation}
\begin{equation}
\label{eq:hom51}
\tilde{r}_N=R_{N,N+1},
\end{equation}
respectively, while (\ref{eq:hom2}) yields
\begin{equation}
\label{eq:hom6}
\tilde{r}_{n}=\frac{\tilde{r}_{n-1}-R_{n-1,n}}{1-\tilde{r}_{n-1}R_{n-1,n}}e^{2kh_{n}}, \quad n=3,4\ldots,N,
\end{equation}
where we used the notation of (\ref{eq:h}) and (\ref{eq:coef}). We can now find $\tilde{r}_2,\tilde{r}_3,\ldots,\tilde{r}_N$ from (\ref{eq:hom5}) and  (\ref{eq:hom6}), and put $\tilde{r}_N$ into (\ref{eq:hom51}) to obtain the eigenvalue equation.

In the simplest case $N=2$, this equation results by equating (\ref{eq:hom5}) and (\ref{eq:hom51}), and is $e^{2kh_2}+R_{12}R_{23}=0$. Therefore, subject to the condition $R=R_{21}R_{23}=-R_{12}R_{23}>1$, there is a unique eigenvalue given by $k=k_0=\frac{\ln{R}}{2h_2}$. In agreement with our analytic-continuation argument (of Section~\ref{sec:sec_vii}), $k_0$  coincides with the real-axis pole in  (\ref{eq:vii2}). It is furthermore seen that the obtained eigenvalue equation has no positive solution when $R=R_{21}R_{23}<1$ so, in this case, there is no solution $V_{n,H}(\rho,z)$ of the postulated form.

For $N=3$, the eigenvalue equation that results from our procedure is
\begin{equation}
\label{eq:hom7}   e^{2k(h_2+h_3)}+R_{12}R_{23}e^{2kh_3}+R_{23}R_{34}e^{2kh_2}+R_{12}R_{34}=0,
\end{equation}
which amounts   to the vanishing of the demonimator $D(k)$ in Appendix~\ref{sec:N-3}. Eqn.~(\ref{eq:hom7})
is more interesting than the one for $N=2$. For instance, when $h_2=h_3$ and $R_{12}=R_{34}=-2R_{23}=3$, there are \textit{two} distinct positive eigenvalues. In this case we have found two linearly independent solutions $V_{n,H}(\rho,z)$. In other words, our general solution is a two-parameter family, which involves two arbitrary real constants.

\section{\label{sec:discussion} Discussion and summary}

This work developed a general framework for the electrostatic analysis of point charges in multilayer planar structures with arbitrary layer thicknesses and material parameters. Starting from the Hankel-transform representation, we derived a systematic formulation for the scalar potential in each layer and showed that the problem can be treated in a compact and physically transparent way through Stokes-like generalized reflection coefficients. In this manner, the classical image-theory picture is preserved, while being extended well beyond the standard parameter regimes in which the corresponding image series is convergent.

A central outcome of the analysis is that the mathematical character of the solution is dictated by the denominator appearing in the transformed-domain representation. For conventional parameter ranges, the relevant integrals converge and lead to unique solutions that admit the familiar interpretation in terms of image charges. The analysis also shows, however, that this standard picture does not persist in all regimes. In particular, when dielectric contrasts of opposite sign are involved, the image series may diverge even though the electrostatic boundary-value problem remains meaningful. This is especially relevant in settings involving negative permittivities, plasmonic resonances, or active effective media, where overscreening effects may arise and where naive image-theory constructions are no longer adequate.

For the three-layer problem and the case of two layers above a ground plane, we found that the parameter $R$ naturally partitions the problem into distinct mathematical regimes. In the CI-CS region, both the defining integral and the associated image series converge, and the conventional image-theory interpretation remains valid. In the CI-DS region, the integral representation remains well defined even though the image series diverges; in this sense, the integral may be viewed as a regularization---obtained via analytic continuation---of the elementary image construction. In the NU-DS region, the transformed-domain integrand presents a simple pole on the positive real axis, leading to non-uniqueness. In that case, analytic continuation in the complex-$R$ plane together with the Plemelj formulas yields a one-parameter family of solutions consisting of a principal-value particular term and a homogeneous contribution.

This non-uniqueness may also be viewed as a sign that the strictly local and static constitutive description is no longer sufficient to identify a unique physical branch. Once poles and homogeneous solutions appear, additional physics (as well as analytic-continuation arguments like those in Section~\ref{sec:sec_vii}) may be needed in order to regularize the problem and select the physically relevant solution. One possible interpretation is through spatial dispersion, since nonlocality can alter the admissible field profiles and regularize singular electrostatic behavior. Another possible interpretation arises in media with active response, because such parameters are generally not compatible with a purely passive static model. From this perspective, the observed non-uniqueness is not merely a mathematical feature of the layered boundary-value problem, but may instead reflect a hidden dependence on constitutive effects beyond the local passive electrostatic approximation.

The obtained homogeneous solutions are of independent interest. For the three-layer problem, they were derived explicitly and connected directly to the pole of the transformed-domain solution. For more layers, we showed that analogous homogeneous solutions can be constructed through an eigenvalue-type procedure, and that multiple positive eigenvalues may occur. This indicates that resonant electrostatic layered systems may possess a richer modal structure than what is apparent from the simplest two- and three-layer examples. Physically, the homogeneous solutions are naturally connected to source-free plasmon-like states supported by the layered geometry.

Another contribution of the paper is the introduction of the phantom-image construction. When $\abs{R}$ is large, the potential can be approximated asymptotically by a finite set of phantom sources located at positions different from those of the standard images. This replaces an inaccessible divergent infinite image sequence by a finite and computationally simple approximation that remains valid in parameter regimes where the traditional image interpretation fails. Although the phantom-image picture is asymptotic and subject to geometric restrictions, it provides an appealing alternative representation of the field and may prove useful for fast estimates and numerical implementations.

The generalized-reflection-coefficient algorithm is also computationally advantageous. Direct treatment of the full $(2N+2)\times(2N+2)$ algebraic system is feasible for small $N$, but becomes cumbersome as the number of layers increases. By contrast, the recursive formulation in terms of generalized reflection coefficients yields a clear layer-by-layer procedure while preserving a close connection to the physics of successive reflections and transmissions across interfaces. This makes the framework attractive not only for symbolic derivations but also for numerical implementations in structures with many layers.

In summary, the paper provides a unified treatment of electrostatic point charges in multilayer planar media that combines Hankel-transform methods, generalized reflection coefficients, principal-value regularization, homogeneous resonant solutions, and asymptotic phantom-image representations. Beyond providing practical computational tools, the analysis clarifies the mathematical structure underlying image theory in layered systems and shows how this classical concept can be extended to regimes in which the standard image-charge series diverges.

\appendix
\section{Expansion of Stokes-like generalized reflection coefficients method for charge located inside the $n'$th layer}
\label{sec:charge-inside}

If the charge $q$ is assumed to be inside the $n'$th layer of the structure: ${d_{n' - 1}} < {d_g} < {d_{n'}}$. Then for $n \ge n'$:

\begin{eqnarray}
{A_n}\left( k \right) = \frac{1}{{{T_{n,n + 1}}}}\left[ {{A_{n + 1}}\left( k \right) + {B_{n + 1}}\left( k \right){R_{n,n + 1}}{e^{2k{d_n}}}} \right],\quad n = n', \ldots ,N
\label{eq:ap1}
 \\
{B_n}\left( k \right) + {\delta _{n'n}}{e^{ - k{d_q}}} = \frac{1}{{{T_{n,n + 1}}}}\left[ {{A_{n + 1}}\left( k \right){R_{n,n + 1}}{e^{ - 2k{d_n}}} + {B_{n + 1}}\left( k \right)} \right],\quad n = n', \ldots ,N
\label{eq:ap2}
\end{eqnarray}

For $n \le n'$:

\begin{eqnarray}
{A_n}(k) + {\delta _{n'n}}{e^{k{d_q}}} = \frac{1}{{{T_{n,n - 1}}}}\left[ {{B_{n - 1}}\left( k \right){R_{n,n - 1}}{e^{2k{d_{n - 1}}}} + {A_{n - 1}}\left( k \right)} \right],\quad n = 1,...,n'
\label{eq:ap3}
 \\
{B_n}\left( k \right) = \frac{1}{{{T_{n,n - 1}}}}\left[ {{B_{n - 1}}\left( k \right) + {A_{n - 1}}\left( k \right){R_{n,n - 1}}{e^{ - 2k{d_{n - 1}}}}} \right],\quad n = 1,...,n'
\label{eq:ap4}
\end{eqnarray}

The problem now breaks down using two groups of generalized coefficients. Namely, for $n \ge n'$, we have

\begin{equation}
{\tilde R_{n,n + 1}}\left( k \right) = \frac{{{A_n}(k){e^{ - 2k{d_n}}}}}{{{B_n}(k) + {\delta _{n'n}}{e^{ - k{d_q}}}}},{\kern 1pt} \quad n = n', \ldots ,N
\label{eq:ap5},
\end{equation}

and for $n \le n'$:

\begin{equation}
{\tilde R_{n,n - 1}}\left( k \right) = \frac{{{B_n}(k){e^{2k{d_{n - 1}}}}}}{{{A_n}(k) + {\delta _{n'n}}{e^{k{d_q}}}}},{\kern 1pt} \quad n = 2,...,n'
\label{eq:ap6},
\end{equation}

For the ${\tilde R_{n,n + 1}}\left( k \right)$, the equations remain the same as in the main text. Eq. (\ref{eq:R2}) of the main text is holds for $n = n', \ldots ,N$. Eq. (\ref{eq:R4}) holds for $n = n', \ldots ,N - 1$. Eq. (\ref{eq:R5}) holds for $n = n', \ldots ,N - 1$ and Eq. (\ref{eq:R7}) holds for $n = n' + 1, \ldots ,N$.

We follow the same steps for ${\tilde R_{n,n - 1}}\left( k \right)$. We get:

\begin{eqnarray}
{\tilde R_{n,n - 1}}\left( k \right) = \frac{{{B_{n - 1}}(k){e^{2k{d_{n - 1}}}} + {A_{n - 1}}(k){R_{n,n - 1}}}}{{{B_{n - 1}}(k){R_{n,n - 1}}{e^{2k{d_{n - 1}}}} + {A_{n - 1}}(k)}},\quad n = 2,...,n'
\label{eq:ap7}
 \\
{\tilde R_{2,1}}(k) = {R_{2,1}}
\label{eq:ap8}
\\
{\tilde R_{n,n - 1}}\left( k \right) = \frac{{{R_{n,n - 1}} + {{\tilde R}_{n - 1,n - 2}}{e^{2k({d_{n - 1}} - {d_{n - 2}})}}}}{{1 + {R_{n,n - 1}}{{\tilde R}_{n - 1,n - 2}}{e^{2k({d_{n - 1}} - {d_{n - 2}})}}}},\quad  n = 3,...,n'
\label{eq:ap9}
\\
{\tilde R_{n,n - 1}}\left( k \right) = {R_{n,n - 1}} + \frac{{{T_{n,n - 1}}{T_{n - 1,n}}{{\tilde R}_{n - 1,n - 2}}{e^{2k({d_{n - 1}} - {d_{n - 2}})}}}}{{1 + {R_{n,n - 1}}{{\tilde R}_{n - 1,n - 2}}{e^{2k({d_{n - 1}} - {d_{n - 2}})}}}},\quad n = 3,...,n'
\label{eq:ap10}
\\
{B_{n + 1}}\left( k \right) = \frac{{{B_n}\left( k \right)}}{{{T_{n + 1,n}}}}\left[ {1 + \frac{{{A_n}\left( k \right)}}{{{B_n}\left( k \right)}}{R_{n + 1,n}}{e^{ - 2k{d_n}}}} \right],\quad n = 2, \ldots ,n' - 1
\label{eq:ap11}
\\
{B_n}(k) = \frac{{{T_{n + 1,n}}{B_{n + 1}}{{\tilde R}_{n,n - 1}}}}{{{{\tilde R}_{n,n - 1}} + {R_{n + 1,n}}{e^{2k({d_{n - 1}} - {d_n})}}}}
\label{eq:ap12}
\end{eqnarray}

The above-derived formulas give rise to the following algorithm for the analytical (and, for many layers, symbolical) calculation of ${A_1}\left( k \right), \ldots ,{A_N}\left( k \right),{B_2}\left( k \right), \ldots ,{B_{N + 1}}\left( k \right)$, with (${B_1}\left( k \right) = {A_{N + 1}}\left( k \right) = 0$).

1) Obtain: ${\tilde R_{N,N + 1}}\left( k \right), \ldots ,{\tilde R_{n',n' + 1}}\left( k \right)$ and ${\tilde R_{2,1}}\left( k \right), \ldots ,{\tilde R_{n',n' - 1}}\left( k \right)$.

2) ${A_{n'}} = {\tilde R_{n',n' + 1}}{\textstyle{{{e^{k(2{d_{n' - 1}} - {d_q})}} + {{\tilde R}_{n',n' - 1}}{e^{k{d_q}}}} \over {{e^{2k({d_{n' - 1}} - {d_{n'}})}} - {{\tilde R}_{n',n' + 1}}{{\tilde R}_{n',n' - 1}}}}}$, ${B_{n'}} = {\tilde R_{n',n' - 1}}{\textstyle{{{e^{ - k(2{d_{n'}} - {d_q})}} + {{\tilde R}_{n',n' + 1}}{e^{ - k{d_q}}}} \over {{e^{2k({d_{n' - 1}} - {d_{n'}})}} - {{\tilde R}_{n',n' + 1}}{{\tilde R}_{n',n' - 1}}}}}$.

3) Determine ${A_{n' + 1}}\left( k \right), \ldots ,{A_N}\left( k \right)$ as a function of ${A_{n'}}$ and ${B_{n' - 1}}\left( k \right), \ldots ,{B_2}\left( k \right)$   as a function of ${B_{n'}}$. 

4) For $n = n' + 1, \ldots ,N$ use the known values of ${A_n}\left( k \right)$ and ${\tilde R_{n,n + 1}}\left( k \right)$ to determine ${B_n}\left( k \right)$. For $n = 2,...,n' - 1$, use the known values of ${B_n}\left( k \right)$ and ${\tilde R_{n,n - 1}}\left( k \right)$ to determine ${A_n}\left( k \right)$.

5) Calculate: ${B_{N + 1}}\left( k \right) = {T_{N,N + 1}}\left( {{B_N}\left( k \right) + {\delta _{n'N}}{e^{ - k{d_g}}}} \right)$, ${A_1}\left( k \right) = {T_{2,1}}\left( {{A_2}\left( k \right) + {\delta _{n'2}}{e^{k{d_g}}}} \right)$. If the charge was located outside the layers, then one of the coefficients ${B_{N + 1}}(k),{A_1}(k)$ is found directly from the generalized reflection coefficients (as seen in the main text).

\section{Some properties of $\Psi$-function}
\label{sec:further-psi}

This appendix proceeds from the definition (\ref{eq:v6}) to derive useful expressions involving ${\Psi}\left( {\rho ,x,\Delta z,R} \right)$, where $x\in\mathbb{R}$ and $\Delta z>0$. For clarity, we often suppress arguments in ${\Psi}$. 

Evidently, $\Psi\Delta z$ is a dimensionless function of three variables (and not four). The three can, for example, be taken to be $\rho/\Delta z$, $\abs{x}/\Delta z$, and $R$; this is apparent from 
\begin{equation}
{\Psi}\left( {\rho ,x,\Delta z,R} \right) = \frac{1}{\Delta z}\int_{0}^\infty J_0\left(\frac{\rho}{\Delta z}u\right)\frac{e^{-\frac{|x|}{\Delta z}u}}{1-Re^{-u}}\,du,\quad R\in \mathbb{C}\setminus [1,+\infty),
\label{eq:app-b0}
\end{equation}
which is an immediate consequence of (\ref{eq:v6}). Setting $u=\ln t$ gives
the Cauchy-type\cite{Fokas1} integral
\begin{equation}
{\Psi}\left( {\rho ,x,\Delta z,R} \right) = \frac{1}{\Delta z}\int_1^\infty  {J_0}\left(\frac{\rho}{\Delta z}\ln{t} \right)\frac{t^{-\frac{\abs{x}}{\Delta z}}}{t-R}\,dt,\quad R\in \mathbb{C}\setminus [1,+\infty).
\label{eq:app-b1}
\end{equation}
Application of the Plemelj formulas\cite{Fokas1} to (\ref{eq:app-b1}) imply that (\ref{eq:app-b1}) also holds on the cut as long as the integral is interpreted as a principal value, see (\ref{eq:psipv-alternative}); and also give the discontinuity across a point $R_0$ of the branch cut:
\begin{equation}
\frac{{\Psi}\left( R_0+i0 \right)-{\Psi}\left( R_0-i0 \right)}{2}= \frac{i\pi}{\Delta z}\,{J_0}\left(\frac{\rho}{\Delta z}\ln{R_0} \right) R_0^{\,-\frac{\abs{x}}{\Delta z}},
\quad R_0\in (1,+\infty).
\label{eq:app-b2}
\end{equation}
Upon setting $k_0=\ln R_0/\Delta z$ and $\beta=i\pi\alpha/\Delta z$ in (\ref{eq:vii302}), we obtain (\ref{eq:vii4}). 

For $R\in\mathbb{C}\setminus [1,+\infty)$, we now derive (\ref{eq:psi7}). In (\ref{eq:app-b0}), replace the $J_0$ via its defining series\cite{NIST}
\begin{equation}
\label{eq:besseldef}
J_0(z)=\sum_{m=0}^\infty \left(-\frac{1}{4}\right)^m\frac{z^{2m}}{(m!)^2},
\end{equation}
and then use (\ref{eq:phiintegralrepresentation}) to integrate term-by-term. This yields (\ref{eq:psi7}) for all $R$ in the cut plane, and  the Plemelj formulas (see (\ref{eq:phiplemelj}) and (\ref{eq:psipv})) then extend (\ref{eq:psi7}) to $R>1$.

For the case $R\in\mathbb{R}$, we now obtain a large-$|R|$ formula for $\Psi$. As $R\to -\infty$ (or $R\to +\infty$), we neglect the $1$ in the denominator appearing in (\ref{eq:v6}) (or in the denominator appearing in (\ref{eq:psipv}), respectively). As long as $\abs{x}>\Delta z$, the resulting integral---which no longer requires a principal value in the case of (\ref{eq:psipv})--- is convergent and can be evaluated with the aid of (\ref{eq:inv2}). The result is the large-$|R|$ asymptotic approximation
\begin{equation}
    {\Psi}\left(\rho,x,\Delta z,R \right) \sim \frac{-1}{R}\frac{1}{\sqrt {{\rho ^2} + {{\left( \abs{x}-\Delta z \right)}^2}}},\textrm{\ as \ }R\to\pm \infty,\quad  \abs{x}-\Delta z>0,
    \label{eq:asymp-2}
\end{equation}
where $\Psi$ stands for $\Psi_\textrm{PV}$ in the case $R\to +\infty$. We stress that (\ref{eq:asymp-2}) is only valid subject to $\abs{x}>\Delta z$.

Assume that $R\in\mathbb{R}\setminus\{1\}$ and apply the identity 
\begin{equation}
    \frac{1}{1-y}=-\sum_{m=-M}^{-1}y^m+\frac{1}{y^M(1-y)},\quad M=0,1,\ldots
    \label{eq:app-b3}
\end{equation}
to (\ref{eq:v6}) or (\ref{eq:psipv}) by setting $R\exp(-k\Delta z)=y$.
Subject to the condition $\abs{x}-M\Delta z>0$, the sum in (\ref{eq:app-b3}) yields a sum of $M$ convergent integrals, and each can be evaluated via (\ref{eq:inv2}); and the other term in (\ref{eq:app-b3}) (i.e., $y^{-M}(1-y)^{-1}$) gives a $\Psi$-function with $\abs{x}-M\Delta z$ in place of $x$. For $R\in\mathbb{R}\setminus\{1\}$ and $M=0,1,2,\ldots$ we have thus obtained
\begin{equation}
    \Psi\left(x,\Delta z\right)=\sum\limits_{m = -M}^{-1}  \frac{-R^m}{{\sqrt {{\rho ^2} + {{\left( {\left| x \right| + m\Delta z} \right)}^2}} }}+\frac{1}{R^{M}}\Psi\left(\abs{x}-M\Delta z,\Delta z\right), \quad \abs{x}-M\Delta z>0,
    \label{eq:app-b4}
\end{equation}
where we retain only the second and third arguments of the two $\Psi$-functions, which are to be understood as $\Psi_\textrm{PV}$ whenever $R>1$. Note that, as long as  $\abs{x}-M\Delta z>0$, the right-hand side of (\ref{eq:app-b4}) is independent of $M$. Note also that the steps leading to (\ref{eq:app-b4}) break down for sufficiently large $M$.

In the case $-1\le R<1$, we can replace the second $\Psi$-function in (\ref{eq:app-b4}) by its image series from (\ref{eq:viii1}). Upon doing so, we see that the right-hand side is simply the image series of the first $\Psi$-function. Thus (\ref{eq:app-b4}) is rather uninteresting when $-1\le R<1$; but when $\abs{R}>1$, it is the foundation for the results in Section~\ref{sec:phantom}.

\section{Four Layers ($N=3$)}
\label{sec:N-3}

For $N=3$, direct application of the algorithm presented in Section~\ref{sec:sec_iii} yields the following coefficients:
\begin{equation}
 {A_1}(k) = \frac{{{e^{ - k{d_q}}}\left[ {{R_{34}}{e^{2k{d_3}}} + {R_{23}}{e^{2k{d_2}}} + {R_{12}}{e^{2k{d_1}}} + {R_{12}}{R_{23}}{R_{34}}{e^{2k({d_3} - {d_2} - {d_1})}}} \right]}}{D(k)},
    \label{eq:app-c1}
\end{equation}
\begin{equation}
{A_2}(k) = \frac{{{T_{12}}{e^{ - k{d_q}}}\left[ {{R_{34}}{e^{2k{d_3}}} + {R_{23}}{e^{2k{d_2}}}} \right]}}{D(k)},
    \label{eq:app-c2}
\end{equation}
\begin{equation}
{B_2}(k) = \frac{{{T_{12}}{e^{ - k{d_q}}}\left[ {1 + {R_{23}}{R_{34}}{e^{2k({d_3} - {d_2})}}} \right]}}{D(k)},
    \label{eq:app-c3}
\end{equation}
\begin{equation}
{A_3}(k) = \frac{{{T_{12}}{T_{23}}{e^{ - k{d_q}}}{R_{34}}{e^{2k{d_3}}}}}{D(k)},
    \label{eq:app-c4}
\end{equation}
\begin{equation}
{B_3}(k) = \frac{{{T_{12}}{T_{23}}{e^{ - k{d_q}}}}}{D(k)},
    \label{eq:app-c5}
\end{equation}
\begin{equation}
{B_4}(k) = \frac{{{T_{12}}{T_{23}}{T_{34}}{e^{ - k{d_q}}}}}{D(k)}, 
    \label{eq:app-c6}
\end{equation}
in which the denominator $D(k)$ is
\begin{equation}
    D(k)={{1 - {R_{21}}{R_{23}}{e^{ - 2k({d_1} - {d_2})}} - {R_{32}}{R_{34}}{e^{ - 2k({d_2} - {d_3})}} - {R_{43}}{R_{12}}{e^{ - 2k({d_1} - {d_3})}}}}.
\end{equation}
The spatial-domain solutions thus involve a generalization $\Xi$ of our $\Psi$-function, viz.,
\begin{equation}
\Xi (\rho ,x,\Delta {z_1},\Delta {z_2},\Delta {z_3},{R_1},{R_2},{R_3}) = \int_0^\infty  {{J_0}(k\rho )\frac{{{e^{ - k\left| x \right|}}}}{{1 - \sum\nolimits_{n = 1}^3 {{R_n}{e^{ - k\Delta {z_n}}}} }}}\,dk.
\end{equation}
Further studies and generalizations may be developed based on this formulation; these will be the subject of subsequent work.

\begin{acknowledgments}
GF thanks Prof. Dionisios Margetis for decisive guidance on all aspects of the non-unique solution, and Dr. Vassilios Houdzoumis for valuable help with the analytic-continuation techniques. TTK acknowledges the support of the Bodossaki Foundation through the ``Stamatis G. Mantzavinos'' Scholarship.
\end{acknowledgments}


%
%

%



\end{document}